\documentclass[11pt,a4paper,english,twoside]{article}

\usepackage{a4wide}
\usepackage{amssymb, amsmath}
\usepackage{graphicx}
\usepackage[all]{xy}
\usepackage{hyperref}
\hypersetup{linktocpage}
\usepackage{enumerate}
\usepackage{xcolor}
\usepackage{dsfont}
\usepackage{empheq}
\usepackage{cite}
\usepackage{float}

\usepackage{soul}

\usepackage{enumitem}

\newcommand{\beq}{\begin{equation}}
\newcommand{\eeq}{\end{equation}}
\def\bea#1\eea{\begin{align}#1\end{align}}
\def\beal#1\eeal{\begin{subequations}\begin{align}#1\end{align}\end{subequations}}
\newcommand{\nn}{\nonumber}
\newcommand{\w}{\wedge}

\def\del {\partial}
\def\d {{\rm d}}
\def\mmm {\mathcal{M}}
\def\tV {\tilde{V}}

\newtheorem{thm}{Conjecture}

\begin{document}
\numberwithin{equation}{section}

\begin{titlepage}

\begin{flushright}
CERN-TH-2019-019
\end{flushright}

\begin{center}

\phantom{DRAFT}

\vspace{2.9cm}

{\LARGE \bf{Open problems on classical de Sitter solutions}}\\

\vspace{2 cm} {\Large David Andriot}\\
 \vspace{0.9 cm} {\small\slshape Theoretical Physics Department, CERN\\
1211 Geneva 23, Switzerland}\\
\vspace{0.5cm} {\upshape\ttfamily david.andriot@cern.ch}\\

\vspace{3cm}

{\bf Abstract}
\vspace{0.1cm}
\end{center}

\begin{quotation}
\noindent Classical 10d string backgrounds with a 4d de Sitter space-time, $D$-brane and orientifold sources, are commonly believed to satisfy the following:
\begin{enumerate}[label=\textbf{\arabic*.}]
\item {\it There is no classical de Sitter solution with parallel sources.}
\item {\it Classical de Sitter solutions with intersecting sources are unstable.}
\item {\it Classical de Sitter solutions cannot have at the same time a large internal volume, a small string coupling, a bounded number of orientifolds and quantized fluxes.}
\end{enumerate}
These three conjectures are of particular relevance to the swampland program, and if true, they challenge the connection of string theory to cosmology. We restrict here to a standard solution ansatz for which the problem is well-defined, and we still fail to prove analytically any of these conjectures. While developing new tools and obtaining new constraints, we identify remaining corners of parameter space where counter-examples to these conjectures could be found.
\end{quotation}

\end{titlepage}

\newpage

\tableofcontents

\section{Introduction}

Relating quantum gravity and cosmology is an important and timely topic. Cosmological observations, both of the early and late universe, have reached unprecedented levels of precision, bringing new and crucial information. On the theory side, many early universe models, and several late universe ones, are valid candidates. It would be helpful to distinguish among them by providing to some of them a quantum gravity origin, or ensuring at least consistency with such a theory: this is at the heart of the on-going swampland program. Such consistency checks revolve around a central question, namely whether the cosmological model, with quantum gravity origin, admits a solution with a de Sitter space-time. This problem matters for the early universe, where inflating space-times can be close to de Sitter, and the post-inflation reheating that typically requires a potential with a de Sitter vacuum, but also for the late universe, described with a cosmological constant and asymptotically de Sitter, or with a quintessence model. We focus here on string theory, and the question becomes whether it can admit a background with a de Sitter space-time; a review on this matter can be found in \cite{Danielsson:2018ztv}. Beyond the cosmological motivation, this is an interesting question per se, with other applications such as holography and the relation to higher spin theories.

In this paper we address this question restricting ourselves to classical perturbative de Sitter string backgrounds, discarding other approaches \cite{Kachru:2003aw, Balasubramanian:2005zx} that involve string loop or non-perturbative contributions, $\alpha'$ corrections, or further ingredients (more recent proposals include \cite{Weissenbacher:2019bfb, Blaback:2019zig, Berglund:2019pxr}). A motivation for our restrictive setting is that it is simpler, the approximations required to reach a final effective theory are a priori easier to control. The limited number of ingredients will allow us to have well-posed mathematical problems, hopefully helping to get some answers. These ingredients are the fields of a 10d theory, here type II supergravities, and extended objects, here $D_p$-branes and orientifold $O_p$-planes. Classical perturbative de Sitter string backgrounds have been excluded from heterotic string \cite{Green:2011cn, Gautason:2012tb, Kutasov:2015eba, Quigley:2015jia} while they remain possible in type II, even though very constrained. So we focus on type II supergravities which are valid low energy, classical and perturbative effective theories of string theory when one can neglect higher $\alpha'$ and string coupling $g_s$ corrections, as we will consider; our framework also discards the F-theory regime. In that case, string backgrounds boil down to solutions of supergravity equations of motion and Bianchi identities, and these solutions are the main focus of this paper. Regarding extended objects, one may still include $N\!S$-sources like $N\!S_5$-branes or $K\!K$-monopoles as recently in \cite{Blaback:2018hdo} (see also \cite{Silverstein:2007ac}), or anti-objects like $\overline{D}_p$-branes as in \cite{Kallosh:2018nrk}. Even though there could be appealing reasons for doing so, we restrict here to $O_p/D_p$ only, with the aim of getting conclusive results for this simple and well-studied setup.

In this framework, we want to investigate such classical de Sitter solutions and their properties, a topic already extensively studied in the past \cite{Maldacena:2000mw, Hertzberg:2007wc, Silverstein:2007ac, Covi:2008ea, Haque:2008jz, Caviezel:2008tf, Flauger:2008ad, Danielsson:2009ff, deCarlos:2009fq, Caviezel:2009tu, Dibitetto:2010rg, Wrase:2010ew, Danielsson:2010bc, Blaback:2010sj, Danielsson:2011au, Shiu:2011zt, Burgess:2011rv, VanRiet:2011yc, Danielsson:2012by, Danielsson:2012et, Gautason:2013zw, Kallosh:2014oja, Junghans:2016uvg, Andriot:2016xvq, Junghans:2016abx, Andriot:2017jhf, Andriot:2018ept, Roupec:2018mbn, Junghans:2018gdb, Banlaki:2018ayh, Cordova:2018dbb, Cribiori:2019clo}. The understanding of these solutions, built from all these works, has led to general beliefs that we formulate below as conjectures. Proving (or disproving) these conjectures has remained so far open problems, on which we report here. The main focus of this literature went to a subset of solutions, specified by an ansatz detailed in section \ref{sec:ansatz}: in short, this ansatz consists in a 4d de Sitter space-time times a 6d compact group manifold, with constant fluxes. The metric and dilaton depend in a specific manner on a single function, the warp factor, as in standard Minkowski solutions, and the warp factor is further set to constant in the ansatz through the ``smearing'' procedure. It is however believed, and checked in some cases, that the specific dependence is such that dilaton and warp factor derivatives cancel each other and drop out of the equations, leaving us with quantities corresponding to the smeared case; this provides a justification for studying an ansatz with smearing. It remains important to keep in mind that this ansatz may not be the only option: by only displaying different warp factors and dilaton, the recent work \cite{Cordova:2018dbb} claims to have obtained a classical de Sitter solution, which would otherwise perfectly fit in our framework. If confirmed, this solution falsifies the following conjecture \ref{conj1}, but it is for now under debate \cite{Cribiori:2019clo}; we will come back to it. In this paper, we stick to the common ansatz of smeared solutions on group manifolds with constant fluxes; (dis)proving the following conjectures for it is a well-defined question. This ansatz has the advantage of allowing the use of 4d theories, namely gauged supergravities, to study these solutions. Indeed, this ansatz corresponds to a Scherk-Schwarz truncation, generally a consistent truncation of 10d supergravity to 4d supergravity, meaning that a solution to the 4d theory is a solution to the 10d one. This map to a 4d problem provides extra tools, e.g.~for the study of stability of solutions: this translates into the analysis of a 4d scalar potential. More general solutions as that of \cite{Cordova:2018dbb} (or e.g.~\cite{Andriot:2015sia} for Minkowski and \cite{Malek:2019ucd} for anti-de Sitter) may not admit such a 4d description, in which case only the 10d investigation is appropriate. We come back in more detail in section \ref{sec:defpb} to these various ans\"atze and the relation to 4d supergravity.

We now present three common conjectures on classical de Sitter solutions as defined above. We express them generally, but have in mind in this paper to (dis)prove them for the ansatz just discussed, i.e.~for smeared solutions on group manifolds with constant fluxes, for which at least the question mathematically well-defined. The conjectures refer to sources, by which we mean $O_p$ and $D_p$. One should distinguish whether they are parallel, meaning along the same dimensions in the 10d space-time, or the opposite case where they intersect.
\begin{thm}
There is no classical de Sitter solution with parallel sources.\label{conj1}
\end{thm}
There is indeed no known example of such solution, except possibly the recent one \cite{Cordova:2018dbb} mentioned above (see also \cite{Cribiori:2019clo}), which does not enter our ansatz. But there is also no proof of this conjecture, despite important constraints on such solutions, summarized e.g.~in the introduction of \cite{Andriot:2018ept}. In the corresponding 4d ${\cal N}=4$ gauged supergravity framework, there is also no known solution, and strong constraints were obtained in \cite{Dibitetto:2010rg}, without fully excluding such solutions either.
\begin{thm}
Classical de Sitter solutions with intersecting sources are unstable. \label{conj2}
\end{thm}
We refer here to a classical, perturbative instability, different than a flat direction or even a metastable case that would be perturbatively stable. In a 4d language, this means that at least one 4d scalar field is at a maximum of its potential, i.e.~tachyonic, implying for the standard slow-roll parameter $\eta_V < 0$. The only known classical de Sitter solutions were obtained on group manifolds in \cite{Caviezel:2008tf, Flauger:2008ad, Danielsson:2009ff, Caviezel:2009tu, Danielsson:2010bc, Danielsson:2011au, Roupec:2018mbn} (except maybe \cite{Cordova:2018dbb}) and they all share the properties that they admit intersecting sources and they are unstable. Despite various ideas, and studies of concrete examples \cite{Covi:2008ea, Danielsson:2011au, Shiu:2011zt, Danielsson:2012by, Danielsson:2012et, Kallosh:2014oja, Junghans:2016uvg, Junghans:2016abx, Andriot:2018ept, Roupec:2018mbn}, no analytic proof of this conjecture has been obtained.
\begin{thm}
Classical de Sitter solutions cannot have at the same time a large internal volume, a small string coupling, a bounded number of orientifolds and quantized fluxes.\label{conj3}
\end{thm}
This conjecture challenges the string origin of the solutions; as supergravity solutions, they remain valid and mathematically well-defined. As explained previously, for them to be trustable classical string background, extra requirements have to be met, and the conjecture implies that no classical de Sitter solution will succeed. The first two requirements are needed to justify the supergravity approximation, i.e.~negligible string states and $\alpha'$ or higher derivative corrections, and string loop corrections. The last two are here for string consistency, i.e.~for orientifolds and fluxes to be string objects and not just supergravity ones; we will come back to this. This conjecture has been recently motivated and studied in \cite{Roupec:2018mbn, Junghans:2018gdb, Banlaki:2018ayh} that considered particular cases or (many) concrete examples, without obtaining a general proof.

These three conjectures are especially relevant in the context of the swampland program. Conjectures \ref{conj1} and \ref{conj3} are in agreement with the conjectured absence of de Sitter solution in effective theories of quantum gravity \cite{Obied:2018sgi} while conjecture \ref{conj2} is along the lines of the refined swampland conjectures \cite{Ooguri:2018wrx} (see also \cite{Andriot:2018wzk, Garg:2018reu, Andriot:2018mav}) that exclude de Sitter minima but allow maxima. The values of $\epsilon_V$ and $\eta_V$, part of the de Sitter swampland conjectures, are however not discussed here, except in section \ref{sec:c} for some recent constraints related to conjecture \ref{conj1}, where we compute the constant $c$ of \cite{Obied:2018sgi}. Proving analytically conjectures \ref{conj1}, \ref{conj2} or \ref{conj3} would then bring credit to the de Sitter swampland conjectures. Beyond this, the three conjectures remain interesting on their own, at least as mathematical questions or for the connection to cosmology.

In this paper, we study conjectures \ref{conj1}, \ref{conj2}, \ref{conj3}, obtain new results and provide new tools, but we fail to prove any of these conjectures. We discuss the reasons for this failure, identify precise cases and corners of parameter space that escape the conjectured behaviors of classical de Sitter solutions. Studying in future work these corners may lead to the opposite result, namely that one finds there counter-examples to these conjectures. This would be equally interesting, and important to build cosmological models. A summary of our results can be found in section \ref{sec:ccl}.

\section{Defining the problem: equations and ansatz}\label{sec:defpb}

In this section, we make the three conjectures \ref{conj1}, \ref{conj2}, \ref{conj3} more precise by providing the appropriate definitions. The main object are ``classical de Sitter solutions'': we will briefly give a general description of those, and then specify in section \ref{sec:ansatz} to the ansatz, i.e.~the class of such solutions, we are interested in; the conjectures are believed to hold at least for this class. Doing so, we take the opportunity to provide some context and literature. We summarize the equations to solve and the ansatz considered in section \ref{sec:sum}. We then make more precise in section \ref{sec:stab} the notion of stability present in conjecture \ref{conj2}. The expert reader may skip this section.\\

We work in 10d type II supergravities with orientifold $O_p$-planes and $D_p$-branes: this framework defines the set of equations that any ``solution'' should solve, namely each equation of motion (e.o.m.) and Bianchi identity (BI):\\
- the 10d Einstein equation\\
- the dilaton $\phi$ e.o.m.\\
- the gauge potentials e.o.m., that we loosely call the $H$-flux e.o.m. and the Ramond-Ramond fluxes $F_q$ e.o.m.\\
- the $H$-flux and $F_q$-fluxes BI\\
- the Riemann tensor BI.\\
We refer to appendix A of \cite{Andriot:2016xvq} for conventions and explicit general equations. In string theory terms, these are closed string background equations, to which one should add open string ones, if $D_p$-branes are present. Here, we discard those from the start, even though they could bring additional interesting constraints in general, related to calibration conditions of $O_p/D_p$. Last but not least, the presence of orientifolds imposes to respect involutions, which will translate into projection conditions on the fields. Given this list of equations or constraints to satisfy, we enter the details of the solutions ansatz.

\subsection{Solution ansatz}\label{sec:ansatz}

\subsubsection*{Ansatz: 0. De Sitter solutions}

We look for de Sitter solutions, by which we mean that the 10d metric is globally written as
\beq
\d s^2_{10}= e^{2A} \d \tilde{s}^2_4 + \d s^2_6 \ ,\label{10dmetric}
\eeq
where $\d \tilde{s}^2_4 = \tilde{g}_{\mu\nu} (x) \d x^\mu \d x^\nu$ is a 4d de Sitter space-time metric, $\d s^2_6= g_{mn} (y) \d y^m \d y^n$ is the metric of a 6d compact manifold $\mmm$, and $e^{A}$ is a function on $\mmm$ called the warp factor. Preserving 4d Lorentz invariance leads to restrictions on the other fields, detailed in \cite{Andriot:2016xvq} as the ``compactification ansatz'': eventually, the only flux degrees of freedom entering our equations are the 3-form $H$ and the $q$-forms $F_q$ on $\mmm$, where $q=0,2,4,6$ in the type IIA theory and $q=1,3,5$ in IIB. Finally, the dilaton is restricted to depend only on 6d (internal) coordinates.

Restrictions are also taken are on the $O_p/D_p$, collectively called the sources. We briefly state those here and refer to \cite{Andriot:2016xvq, Andriot:2017jhf} for more detail. We choose their size to be $ p \geq 3$ and to stand along (at least) the three space dimensions of de Sitter, preserving 4d Lorentz invariance. We take BPS sources, in the sense that their charge is related to their tension $\mu_p= T_p$, and we consider a vanishing pull-back for the world-volume fields $-\iota^*[b] + {\cal F} =0$. Also, $\alpha'$ corrections to their world-volume action, as considered in \cite{Dasgupta:1999ss}, are here not included. Finally, even though a classical de Sitter solution was found with both $O_5$ and $O_7$ in \cite{Caviezel:2009tu}, we consider here sources of a single size $p$; the generalization of our formalism to multiple sizes was considered in \cite{Andriot:2017jhf}. We present below a last, important, restriction on the sources, related to the internal geometry.

\subsubsection*{Ansatz: 1. Sources, warp factor and dilaton}

The restrictions detailed so far match standard compactification ans\"atze; we now specify more. To clarify directions in $\mmm$ that are parallel or transverse to each source, i.e.~its embedding, we use the 6d flat (orthonormal) basis with metric $\delta_{ab}=e^m{}_a e^n{}_b g_{mn}$ and 1-forms $e^a =e^a{}_m \d y^m$ in terms of vielbeins. We assume that for each source, these 1-forms split into two sets $\{ e^{a_{||}} \}$ and $\{ e^{a_{\bot}} \}$, that are globally distinct. This means that the structure group of the cotangent bundle is reduced from $O(6)$ to $O(p-3) \times O(9-p)$, or a subgroup thereof. Note that no assumption is made on the coordinate dependence. This is actually a rather generic situation as further discussed in \cite{Andriot:2016xvq}. We now take the following ansatz: parallel or transverse directions are scaled with the warp factor as $e^a=e^{\pm A} \tilde{e}^{a}$, i.e.~the ansatz for the 10d metric \eqref{10dmetric} becomes
\bea
& \d s^2_{10}= e^{2A} ( \d \tilde{s}^2_{4} + \d \tilde{s}^2_{6||} ) + e^{-2A} \d \tilde{s}^2_{6\bot} \ ,\ {\rm where}\ \d \tilde{s}^2_{6||}= \delta_{ab} \tilde{e}^{a_{||}} \tilde{e}^{b_{||}} \ ,\ \d \tilde{s}^2_{6\bot}= \delta_{ab} \tilde{e}^{a_{\bot}} \tilde{e}^{b_{\bot}} \ ,\label{metricwarped}\\
& \forall\, a_{||},\ e^m{}_{a_{||}} \del_m A = \del_{a_{||}} A = e^{-A} \del_{\tilde{a}_{||}} A = 0 \ ,
\eea
where the warp factor dependence is restricted to transverse directions only. In the case of intersecting sources, the warp factor is typically set to $1$ as we will see, but the metric is still required to decompose as above for each set of parallel sources; this implies a further reduction of the structure group. Our metric ansatz \eqref{metricwarped} goes together with one for the dilaton, fixed in terms of a constant $g_s$ as follows
\beq
e^{\phi} = g_s\, e^{A (p-3)} \ .\label{dil}
\eeq
The ansatz \eqref{metricwarped} and \eqref{dil} is the one defining brane solutions in Minkowski space-time, and was used to find classical Minkowski solutions of type II supergravities \cite{Giddings:2001yu, Grana:2006kf, Andriot:2016ufg}. It was argued that under certain circumstances (e.g.~supersymmetry) \cite{Grana:2006kf, Blaback:2010sj}, one could first find ``smeared'' solutions (or ``in the smeared limit'') where $A=0$, and then obtain a full solution by reintroducing the warp factor with the above scaling (together with an additional flux component called below $F_k^{(0)}$ and adapting the source contribution $T_{10}$). Such a procedure would be justified in general if all derivatives of $A$ and $\phi$ drop out of the equations to be solved when using the above ansatz, leaving only smeared, i.e.~tilded, quantities. This holds for a standard class of Minkowski solutions \cite{Andriot:2016ufg}; it was also verified for the few equations of motion used in \cite{Andriot:2016xvq} to derive no-go theorems on classical de Sitter solutions. The idea that such a procedure should work in general is implicit in many studies of de Sitter solutions, which are then only pursued in the smeared limit. This also motivates us to consider a smeared ansatz in the following (see ansatz: 2).

Even though the ansatz \eqref{metricwarped}, \eqref{dil} is commonly used and well-motivated, it may be too restrictive. First, fixing this way the dilaton forbids to consider F-theory solutions, thus restricts to perturbative solutions. In addition, the first classical de Sitter solution with parallel sources was claimed to be found recently in \cite{Cordova:2018dbb}, partially numerically, by considering a different warping ansatz in the metric, independent of the dilaton; see however \cite{Cribiori:2019clo} for some criticism. This solution and ansatz follow a serie of works where new solutions were identified on Minkowski \cite{Macpherson:2016xwk} or anti-de Sitter \cite{Cordova:2018eba}, with the same idea of having several independent functions for the metric and the dilaton. It is true that our ansatz \eqref{metricwarped}, \eqref{dil} is inspired by Minkowski solutions, and deviations could be considered on different 4d space-times: this is reminiscent of $D_p$ calibration that differs between Minkowski \cite{Koerber:2005qi, Martucci:2005ht} and anti-de Sitter \cite{Koerber:2007jb}. These ideas may open new avenues to find de Sitter solutions, but they have no impact on the initial question of this paper, that is to (dis)prove the three conjectures for the ansatz \eqref{metricwarped}, \eqref{dil}, or the smeared version, that we now turn to.

\subsubsection*{Ansatz: 2. Smearing, group manifolds and constant fluxes}

We consider from now on the smeared version of the previous ansatz, where we set $A=0$ and drop all tilde on the various quantities. One needs further to drop the flux component $F_k^{(0)}$, usually expected to be given by a derivative of the warp factor; it will be set later to zero due to the orientifold projection on constant fluxes. One also needs to consider the integrated value of the source contributions, that trades the localized $\delta$-functions for integers. A motivation for restricting to this simpler ansatz was explained previously: it is believed that non-smeared or localized solutions can be obtained by undoing the above and simply rescaling quantities with the warp factor. Another reason for a smeared ansatz is the consideration of intersecting sources. This case is parameterized \cite{Andriot:2017jhf} by several sets $I=1 \dots N$ of parallel $O_p/D_p$ that intersect each other, i.e.~the pairs of sets $\{\{ e^{a_{||_I}} \}, \{ e^{a_{\bot_I}} \}\}$ and $\{\{ e^{a_{||_J}} \}, \{ e^{a_{\bot_J}} \}\}$ are different for $I\neq J$: they do not wrap (entirely) the same submanifolds, contrary to parallel sources. For intersecting sources, one typically considers a smeared ansatz. It is a standard supergravity problem to find solutions with fully localized intersecting sources, and string theory is believed to complete the solution; we refer to \cite{Andriot:2017jhf, Andriot:2018wzk} for more discussion. In any case, whether these completions of solutions from smeared to localize are valid does not affect the question tackled in this paper: (dis)proving the conjectures for the smeared ansatz.

All known classical de Sitter solutions (with intersecting sources) were found on $\mmm$ being a group manifold, see \cite{Danielsson:2011au} and \cite{Caviezel:2009tu}: we then restrict to such manifolds. These are built from underlying Lie algebras with structure constants $f^a{}_{bc}$. Thanks to Maurer-Cartan equations, the $f^a{}_{bc}$ correspond to spin connection coefficients, so the Ricci tensor can be expressed in terms of them. More precisely, the 1-forms defined previously verify $\d e^a= -\tfrac{1}{2} f^{a}{}_{bc} e^b\w e^c$, i.e.~$f^{a}{}_{bc} = 2 e^a{}_m \del_{[b} e^m{}_{c]}$, and one obtains the Ricci tensor
\beq
2\ {\cal R}_{cd} = - f^b{}_{ac} f^a{}_{bd} - \delta^{bg} \delta_{ah} f^h{}_{gc} f^a{}_{bd} + \frac{1}{2} \delta^{ah}\delta^{bj}\delta_{ci}\delta_{dg} f^i{}_{aj} f^g{}_{hb} \ , \label{Ricci}
\eeq
where the compactness of $\mmm$ was used, giving the sum $f^a{}_{ab}=0$. With a smeared ansatz and without other sources than the $O_p/D_p$, the Riemann tensor BI on a group manifold boils down to the Jacobi identity of the Lie algebra (see e.g.~equation (3.5) of \cite{Andriot:2014uda}), and is then automatically satisfied.

Thanks to the smeared ansatz and the restriction to group manifolds, the Ricci tensor and connections are constant. Through the various equations to solve, it becomes natural to consider the remaining fields as constant: we thus take as an ansatz constant fluxes. By this we mean that the component coefficients of fluxes in the flat basis are constant; this will lead to dramatic simplifications. Let us first introduce the following notations: for any internal $q$-form $F_q$, we denote by a label ${}^{(n)}$ its number of legs along a source with $0 \leq n \leq p-3$, meaning
\beq
F_q = \frac{1}{q!} F^{(0)}_{a_{1\bot} \dots a_{q\bot}} e^{a_{1\bot}} \w \dots \w e^{a_{q\bot}} + \frac{1}{(q-1)!} F^{(1)}_{a_{1||} a_{2\bot} \dots a_{q\bot}} e^{a_{1||}} \w e^{a_{2\bot}} \w \dots \w e^{a_{q\bot}} + \dots \label{fluxes}
\eeq
where now each $F^{(n)}_{q\, a_1 \dots a_q}$ is constant. For intersecting sources, the notion of parallel and transverse depends on the set $I$ so we label all indices with $I$, e.g.~$F_q^{(n)_I}$.

For $p=4,5,6$, the orientifold projection only allows the following fluxes \cite{Andriot:2018ept}
\bea
& O_6: \quad F_0^{(0)}, F_2^{(1)}, F_4^{(2)}, F_6^{(3)},\quad H^{(0)}, H^{(2)} \nn\\
& O_5: \quad F_1^{(0)}, F_3^{(1)}, F_5^{(2)}, \phantom{\ F_6^{(3)}} \quad H^{(0)}, H^{(2)} \label{fluxcompo}\\
& O_4: \quad F_2^{(0)}, F_4^{(1)}, \phantom{\ F_4^{(2)}, F_6^{(3)}} \quad H^{(0)} \nn
\eea
With $k=8-p$, $F_{k}$ being the flux sourced by the $D_p/O_p$, the non-zero flux components are
\beq
F_{k-2}^{(0)},\ F_k^{(1)},\ F_{k+2}^{(2)},\ F_{k+4}^{(3)} \ , \label{Fkns}
\eeq
with $F_{q\geq 7}=0$ and $F_q^{(n> p-3 )}=0$. In addition, the orientifold needs to be compatible with the geometry, which translates into the following structure constants being allowed
\beq
f^{a_{||}}{}_{b_{\bot}c_{\bot}},\ f^{a_{\bot}}{}_{b_{\bot}c_{||}},\ f^{a_{||}}{}_{b_{||}c_{||}} \ .\label{fabcOp}
\eeq
Those also get indices $I$ in case of intersecting sources. The Ricci tensor and scalar simplify and get expressed only in terms of these structure constants: see e.g.~\eqref{R6group}. This reduction of the number of degrees of freedom is very compelling.

\subsection{Equations and ansatz: summary}\label{sec:sum}

We have motivated a simple, though common, ansatz of de Sitter solution of type II supergravities with $O_p/D_p$, that we now summarize. In short, we look for de Sitter solutions with smeared sources on group manifolds with constant fluxes; the $O_p/D_p$ sources can be parallel or intersecting. In more details, the 10d metric is given by \eqref{metricwarped} and the dilaton by \eqref{dil}, with $A=0$ and dropping the tilde notation. The 6d compact manifold is a group manifold based on a Lie algebra with structure constants $f^a{}_{bc}$; the Ricci tensor is given in \eqref{Ricci}. The fluxes are given in \eqref{fluxes} with constant coefficients. The orientifold projection then reduces the degrees of freedom, such that the only variables entering the equations are the structure constants \eqref{fabcOp}, the flux components \eqref{fluxcompo}, and the source contributions given by the constant numbers $T_{10}^I$ and $T_{10} = \sum_{I} T_{10}^I$ introduced in appendix \ref{ap:eq}. In the case of intersecting sources, overlap numbers will also appear, but will then be fixed here.

With this ansatz, the orientifold projection and the Riemann tensor BI are automatically satisfied. The remaining constraints and equations to solve (see the beginning of section \ref{sec:defpb}) become the Einstein equation and the dilaton e.o.m. detailed in appendix \ref{ap:eq}, the fluxes e.o.m. given here by
\bea
& \d( *_6 F_q ) + H \w *_6 F_{q+2} = 0\ \ (1 \leq q \leq 4) \ ,\\
& \d ( *_6 H) - g_s^2 \sum_{0\leq q \leq 4} F_{q} \w *_6 F_{q+2} = 0 \ ,
\eea
and the fluxes BI detailed in appendix A of \cite{Andriot:2016xvq}, that include $\d H =0$ and the one for the sourced flux $F_k$, $ 0 \leq k=8-p \leq 5$
\beq
\d F_k - H \w F_{k-2} = \varepsilon_p \, \sum_{I} \frac{T_{10}^I}{p+1} \, {\rm vol}_{\bot_I}  \ , \label{BI2}
\eeq
with $F_{-1} =F_{-2}=0$ and $\varepsilon_p=(-1)^{p+1} (-1)^{\left[\frac{9-p}{2} \right]}$. This BI can be projected on each volume form ${\rm vol}_{\bot_I}$ transverse to the set $I$, or along different directions.

This finite set of degrees of freedom and equations makes the problem of finding solutions mathematically well-defined. It also suggests a numerical treatment, as done e.g.~in \cite{Danielsson:2011au, Roupec:2018mbn}. Regarding analytical studies of solution properties, the task remains involved as we will see, because most equations are quadratic in the variables.

\subsection{4d supergravities, scalar fields and stability}\label{sec:stab}

As mentioned in the introduction, an alternative way to find solutions is to solve equations of a 4d effective theory. This approach is valid if the 4d solutions can be promoted to 10d ones: this defines a ``consistent truncation''. The solution ansatz just described typically realises a consistent truncation, known as a Scherk-Schwarz truncation \cite{Scherk:1979zr}: the solutions we are interested in can be obtained in a 4d theory that is a gauged supergravity \cite{Andrianopoli:2005jv, Villadoro:2005cu, DallAgata:2009wsi, Dibitetto:2010rg}. Important restrictions should however be imposed on the allowed gaugings in those theories, to ensure a 10d origin on a compact $\mmm$ with orientifold \cite{DallAgata:2009wsi, Dibitetto:2010rg, Dibitetto:2011gm}; this discards for instance de Sitter solutions as those of \cite{deRoo:2002jf} and references therein. Group manifolds being parallelizable, supersymmetry only gets broken by the sources: parallel sources lead to a 4d ${\cal N}=4$ gauged supergravity, while the intersection of $N=4$ $O_6/D_6$ of the kind considered e.g.~in \cite{Danielsson:2011au} gives ${\cal N}=1$. In these 4d theories, the main degrees of freedom are scalar fields $\varphi$ whose dynamics are captured by a scalar potential $V(\varphi)$. We are interested in solutions with static scalars, so the e.o.m. are given by $\del_{\varphi} V = 0$. By definition, these equations are a subset, or equivalent to those listed previously. Such 4d theories have been used to find de Sitter solutions fitting the above ansatz: no such de Sitter solution is known in 4d ${\cal N}=4$ gauged supergravities, and some were found in an ${\cal N}=1$ theory, e.g.~in \cite{Danielsson:2011au}.

Conjecture \ref{conj2} is about the stability of solutions, i.e.~their behavior under fluctuations. In the 4d language, the fluctuations are captured by the scalar fields, and the stability is studied by looking at the Hessian of the potential, or more precisely at the slow-roll parameter $\eta_V$: a maximum along a direction in the scalar field space gives a negative eigenvalue of the Hessian, and $\eta_V < 0$. The solution is then unstable, the corresponding scalar field is tachyonic. When using 4d gauged supergravities and finding a concrete solution, the stability can easily be tested. Finding general constraints on stable de Sitter solutions is more difficult, some were still obtained for some specific 4d ${\cal N}=1$ gauged supergravity e.g.~in \cite{Danielsson:2012by}.

To prove conjectures \ref{conj1}, \ref{conj2}, \ref{conj3}, an in-between setting has been considered following \cite{Hertzberg:2007wc, Silverstein:2007ac, Danielsson:2012et}: one only studies a subset of the scalar fields, namely the ``volume'' $\rho$, the ``dilaton'' $\tau$, and another metric fluctuation $\sigma$ distinguishing internal ${\rm vol}_{||}$ and ${\rm vol}_{\bot}$. Considering the first and second derivatives of the potential with respect to those already sets important constraints, leaving only few cases where the conjectures could still be wrong. We come back to this method in the next sections.

\section{Conjecture 1: no solution with parallel sources}\label{sec:conj1}

Many constraints have been obtained against the existence of classical de Sitter solutions with parallel sources: a summary of those can be found in the introduction of \cite{Andriot:2018ept}. The only possibilities left, preventing us from a proof of conjecture \ref{conj1}, are the cases $p=4,5,6$. For each of those, analogous ingredients are required for the existence: one needs $T_{10}>0$, ${\cal R}_6 < 0$, $F_{k-2} \neq 0$, as well as some non-zero $f^{a_{||}}{}_{b_{\bot}c_{\bot}}$ and $f^{a_{\bot}}{}_{b_{\bot}c_{||}}$. Additional constraints apply to these various quantities. So a large part of parameter space forbids de Sitter solutions, and only a remaining small part, having these appropriate ingredients, may still allow for it. To reach this conclusion, only few of all equations listed in section \ref{sec:sum} have been used: the 4d and 10d traces of Einstein equation, the dilaton e.o.m., the trace of the Einstein equation along internal directions parallel to the source, the $F_k$ BI projected on ${\rm vol}_{\bot}$, the $F_{k-4}$ BI (only non-trivial for $p=4$). The first four equations correspond in 4d to the 4d Einstein equation and $\del_{\rho} V = \del_{\tau} V = \del_{\sigma} V =0$. To complete the proof of conjecture \ref{conj1}, one therefore needs the information or constraints brought by the remaining equations: this may allow to exclude the remaining part of parameter space. We discuss this idea in section \ref{sec:othereqpar}; before we briefly come back to known constraints and their relation to swampland conjectures.

\subsection{Aparte: computing $c$ for recent no-go theorems}\label{sec:c}

The de Sitter swampland conjectures of \cite{Obied:2018sgi, Ooguri:2018wrx} propose the inequality
\beq
|\nabla V| \geq c\, V \ ,
\eeq
where we set the 4d Planck mass $M_p=1$ and $c >0$ is a constant, conjectured to be of order $1$. One defines $|\nabla V| = \sqrt{g^{ij} \del_{\phi^i} V \del_{\phi^j} V}$ where  $g_{ij}(\phi^k)$ is the field space metric, and $\epsilon_V = \frac{1}{2} \left(\tfrac{|\nabla V|}{V}\right)^2$. For $V>0$, the above inequality can be rewritten
\beq
\sqrt{2 \epsilon_V} \geq c \ .
\eeq
This inequality implies a no-go theorem on de Sitter solutions, since at a de Sitter critical point, $|\nabla V| = 0$, so the cosmological constant related to $V$ at this point should be negative or zero. Similarly, a no-go theorem on de Sitter solutions can be obtained by deriving an inequality on a linear combination of $V$ and its first derivatives. In that case, there is a method to compute a corresponding constant $c$ \cite{Hertzberg:2007wc}. This was used in \cite{Obied:2018sgi} for various no-go theorems previously derived against classical de Sitter solutions with parallel sources, to verify that $c \simeq O(1)$. Here, we would like to do the same for new no-go theorems derived in \cite{Andriot:2018ept}: those also take the form of a linear combination, so they are appropriate to this exercise. They depend on the parameter
\beq
\lambda= -\frac{\delta^{cd}   f^{b_{\bot}}{}_{a_{||}  c_{\bot}} f^{a_{||}}{}_{ b_{\bot} d_{\bot}} }{\tfrac{1}{2} \delta^{ab} \delta^{cd} \delta_{i j}  f^{i_{||}}{}_{a_{\bot} c_{\bot}} f^{j_{||}}{}_{b_{\bot} d_{\bot}}} \ ,\label{lambda}
\eeq
and are expressed by the two following inequalities, with $A=p-9, B=p-3$,
\bea
\lambda\leq 0:&\quad 2 V + \frac{3}{2} \ \tau \del_{\tau} V + \frac{A+B}{A-B} \ \rho \del_{\rho} V  + \frac{2}{B-A} \ \sigma \del_{\sigma} V \leq 0 \ , \label{l<0}\\
\lambda\geq 1:&\quad 2 V + \frac{1}{2} \frac{A-5B}{A-3B} \ \tau \del_{\tau} V  + \frac{1}{3} \ \rho \del_{\rho} V  + \frac{2}{3(A-3B)} \ \sigma \del_{\sigma} V \leq 0 \ , \label{l>1}
\eea
that should hold for any solution.

The method to compute $c$ goes as follows \cite{Hertzberg:2007wc}: one has
\beq
\epsilon_V = \frac{1}{2} \sum_i \left(\frac{\del \ln V}{\del \hat{\phi}^i} \right)^2 \geq \frac{1}{2} \left( \left(\frac{\del \ln V}{\del \hat{\rho}} \right)^2 + \left(\frac{\del \ln V}{\del \hat{\tau}} \right)^2 + \left(\frac{\del \ln V}{\del \hat{\sigma}} \right)^2 \right) \label{ineqeps}
\eeq
where $\hat{\phi}^i$ is the canonically normalized field. This quantity is actually measuring a distance to the origin in a space of coordinates $\{ \del_{\hat{\phi}^i} \ln V \}$. Now consider a no-go theorem of the form
\beq
a V + \sum_i b_i \del_{\hat{\phi}^i} V \leq 0 \ ,\ a>0\ ,\ \exists\, b_i\neq 0 \ .\label{nogo}
\eeq
This selects half of the space of coordinates $\{ \del_{\hat{\phi}^i} \ln V \}$: the allowed region is on one side of the hyperplane defined by picking the equality in \eqref{nogo}. For $V>0$ and because $a>0$, the allowed region does not include the origin. Therefore the sphere centered on the origin admits a minimal radius, reached at the point on the hyperplane, closest to the origin. This minimal radius, corresponding to the minimum of $\epsilon_V$, is computed by plugging the hyperplane equation into $\epsilon_V$ and extremising the corresponding function. The reasoning also works for the subset of scalar fields involved in the no-go, using an inequality w.r.t. $\epsilon_V$ like \eqref{ineqeps}. So we compute the minimum of the following function, and find
\beq
f(x,y)= \left(-\frac{a}{b_1}-\frac{b_2}{b_1}x - \frac{b_3}{b_1}y \right)^2 + x^2 + y^2 \ ,\quad {\rm Min}(f)= \frac{a^2}{b_1^2 +b_2^2 + b_3^2} \ .
\eeq
The result is independent of which $b_i\neq0$ has been chosen. We deduce
\beq
2 \epsilon_V \geq \frac{a^2}{b_{\hat{\rho}}^2 + b_{\hat{\tau}}^2 + b_{\hat{\sigma}}^2 } = c^2 \ .
\eeq

To determine $c$, we now put the no-go theorems \eqref{l<0} and \eqref{l>1} in the form \eqref{nogo}, i.e.~define the canonically normalized fields. To that end, we determine the kinetic term of $\sigma$ in appendix \ref{ap:kin}, building on \cite{Roest:2004pk, Hertzberg:2007wc}: we obtain (without Planck mass here)
\beq
\hat{\rho} = \sqrt{\frac{3}{2}} \ln \rho \ , \ \hat{\tau} = \sqrt{2} \ln \tau \ , \ \hat{\sigma} = \sqrt{\frac{-3AB}{2}} \ln \sigma \ . \label{canfields}
\eeq
We deduce the following $c$ constants for the no-go theorems of \cite{Andriot:2018ept} expressed in \eqref{l<0}, \eqref{l>1}
\bea
\lambda \leq 0:&\quad\quad c^2 = \frac{8 (A-B)^2}{9 (A-B)^2 + 3 (A+B)^2 - 12 AB } = \frac{2}{3} \ ,\\
\lambda \geq 1:&\quad\quad c^2 = \frac{24 (A-3B)^2}{3 (A-5B)^2 + (A-3B)^2 - 4  AB } = 6\, \frac{A-3B}{A-7B} \ ,
\eea
meaning for $p=4,5,6$
\beq
\lambda \leq 0:\ \, c \simeq 0.82 \ ;\quad \quad \lambda \geq 1:\ \, c_{p=6} \simeq 1.73,\ c_{p=5} \simeq 1.83,\ c_{p=4} = 2 \ .
\eeq
The proposal $c \simeq O(1)$ of \cite{Obied:2018sgi} is verified.

\subsection{More constraints from the remaining equations}\label{sec:othereqpar}

As discussed at the beginning of section \ref{sec:conj1}, proving conjecture \ref{conj1} and discarding remaining possibilities in parameter space requires extra information from equations not considered so far. To start with, one may consider components of the Einstein equation, instead of the traces. The 4d components derived in appendix \ref{ap:eq} are equivalent to the 4d trace, which has been used already. So we turn to internal components, and restrict to the ansatz specified in section \ref{sec:ansatz}, namely smeared (parallel) sources with an orientifold, on a compact group manifold with constant fluxes. Internal components of the general Einstein equation given in appendix \ref{ap:eq} then become, in flat indices
\bea
{\cal R}_{ab} & = \frac{1}{4} H_{acd}H_b^{\ \ cd}+\frac{g_s^2}{2}\left(F_{2\ ac}F_{2\ b}^{\ \ \ \ c} +\frac{1}{3!} F_{4\ acde}F_{4\ b}^{\ \ \ cde} \right) \\
& + \frac{g_s}{2}T_{ab} + \frac{\delta_{ab}}{16} \left( - g_s T_{10} - 2|H|^2 + g_s^2(|F_0|^2 - |F_2|^2 -3 |F_4|^2 + 3 |F_6|^2 ) \right) \nn \ ,\\
{\cal R}_{ab} & = \frac{1}{4} H_{acd}H_b^{\ \ cd}+\frac{g_s^2}{2}\left(F_{1\ a}F_{1\ b} +\frac{1}{2!} F_{3\ acd}F_{3\ b}^{\ \ \ cd} + \frac{1}{2 \cdot 4!} F_{5\ acdef}F_{5\ b}^{\ \ \ cdef} - \frac{1}{2} *_6 F_{5\ a} *_6 F_{5\ b} \right) \nn\\
& + \frac{g_s}{2}T_{ab} + \frac{\delta_{ab}}{16} \left( - g_s T_{10} - 2|H|^2 - 2 g_s^2 |F_3|^2 \right) \nn \ .
\eea
There are three cases to consider, with respect to the sources: parallel components ${}_{a_{||}b_{||}}$, transverse ones ${}_{a_{\bot}b_{\bot}}$, and ``off-diagonal'' ones ${}_{a_{||}b_{\bot}}$.

We start with the off-diagonal ${}_{a_{||}b_{\bot}}$ components of the internal Einstein equation. The source term is $T_{ab} = \delta^{a_{||}}_{a} \delta^{b_{||}}_{b} \delta_{a_{||}b_{||}} \frac{T_{10}}{p+1}$, so it does not contribute. The same holds for the Ricci tensor given in \eqref{Ricci}: with the structure constants given in \eqref{fabcOp}, one verifies that ${\cal R}_{c_{||}d_{\bot}} = 0 $. Finally, fluxes contributions also vanish. The $H$-flux only has $H^{(0)}$ or $H^{(2)}$ components, i.e.~no or two legs along parallel directions, so the contraction $H_{a_{||} cd}H_{b_{\bot}}^{\ \ cd}$ is forced to vanish. The same goes for $p=4,5,6$ for the allowed components of the RR fluxes \eqref{Fkns}: contractions $ F_{2\ a_{||}c}F_{2\ b_{\bot}}^{\ \ \ \ c},\ F_{4\ a_{||}cde}F_{4\ b_{\bot}}^{\ \ \ cde}$ and $F_{3\ a_{||}cd}F_{3\ b_{\bot}}^{\ \ \ cd}, \ F_{5\ a_{||}cdef}F_{5\ b_{\bot}}^{\ \ \ cdef}$ are forced to vanish. In addition, $F_1$ and $*_6 F_{5}$ are purely transverse for $p=5$, so their ${}_{a_{||}b_{\bot}}$ contribution vanishes. We conclude that these off-diagonal Einstein equations are automatically satisfied.

We turn to the parallel components ${}_{a_{||}b_{||}}$. The equations become (for $p=4,5,6$)
\bea
{\cal R}_{a_{||}b_{||}} & = \frac{1}{4} H^{(2)}_{a_{||}cd}H_{b_{||}}^{(2) \ cd}+\frac{g_s^2}{2}\left(F_{2\ a_{||}c}F_{2\ b_{||}}^{\ \ \ \ c} +\frac{1}{3!} F_{4\ a_{||}cde}F_{4\ b_{||}}^{\ \ \ cde} \right) \\
& + \frac{\delta_{a_{||}b_{||}}}{16} \left( g_s \frac{T_{10}}{p+1} (7-p) - 2|H|^2 + g_s^2(|F_0|^2 - |F_2|^2 -3 |F_4|^2 + 3 |F_6|^2 ) \right) \nn \ ,\\
{\cal R}_{a_{||}b_{||}} & = \frac{1}{4} H^{(2)}_{a_{||}cd}H_{b_{||}}^{(2) \ cd}+\frac{g_s^2}{2}\left(\frac{1}{2!} F_{3\ a_{||}cd}F_{3\ b_{||}}^{\ \ \ cd} + \frac{1}{2 \cdot 4!} F_{5\ a_{||}cdef}F_{5\ b_{||}}^{\ \ \ cdef} \right) \nn\\
& + \frac{\delta_{a_{||}b_{||}}}{16} \left(  g_s \frac{T_{10}}{p+1} (7-p) - 2|H|^2 - 2 g_s^2 |F_3|^2 \right) \nn \ .
\eea
We can replace the bracket quantity using the 4d Einstein equation equivalent to
\beq
4 {\cal R}_4 =  - 2 |H|^2 + g_s (7-p) \frac{T_{10}}{p+1} + g_s^2 \sum_{q=0}^{6} (1-q) |F_q|^2 \ .
\eeq
We deduce
\bea
{\cal R}_{a_{||}b_{||}} & = \frac{1}{4} H^{(2)}_{a_{||}cd}H_{b_{||}}^{(2) \ cd}+\frac{g_s^2}{2}\left(F_{2\ a_{||}c}F_{2\ b_{||}}^{\ \ \ \ c} +\frac{1}{3!} F_{4\ a_{||}cde}F_{4\ b_{||}}^{\ \ \ cde} \right) + \frac{\delta_{a_{||}b_{||}}}{4} \left( {\cal R}_4 + 2 g_s^2 |F_6|^2  \right) \nn \ ,\\
{\cal R}_{a_{||}b_{||}} & = \frac{1}{4} H^{(2)}_{a_{||}cd}H_{b_{||}}^{(2) \ cd}+\frac{g_s^2}{2}\left(\frac{1}{2!} F_{3\ a_{||}cd}F_{3\ b_{||}}^{\ \ \ cd} + \frac{1}{2 \cdot 4!} F_{5\ a_{||}cdef}F_{5\ b_{||}}^{\ \ \ cdef} \right) + \frac{\delta_{a_{||}b_{||}}}{4} \left( {\cal R}_4 + g_s^2 |F_5|^2 \right) \nn \ .
\eea
For diagonal components, namely $a_{||}=b_{||}$, each term in the right-hand side is positive: indeed, flux contractions are then sums of squares. We deduce that a de Sitter solution requires
\beq
\mbox{de Sitter}:\quad \forall\ a_{||}\ ,\ {\cal R}_{a_{||}a_{||}} > 0 \ .\label{reqRpar}
\eeq
We actually knew already this property for the trace of the above \cite{Andriot:2016xvq}, providing important constraints; we now have a stronger version on each diagonal component of the Ricci tensor.\footnote{The property \eqref{reqRpar}, or even that of the trace derived with warp factor and dilaton (ansatz 1) in \cite{Andriot:2016xvq}, is not satisfied by the solution of \cite{Cordova:2018dbb}: there, an Einstein manifold with negative curvature is considered along the source. If valid \cite{Cribiori:2019clo}, the different ansatz of that solution discussed in section \ref{sec:ansatz} then leads to important differences in the allowed internal geometries.} Looking back at the Ricci tensor expression \eqref{Ricci}, the first two terms can be written as minus a square
\beq
 - f^b{}_{ac_{||}} f^a{}_{bc_{||}} - \delta^{bg} \delta_{ah} f^h{}_{gc_{||}} f^a{}_{bc_{||}} = - \frac{1}{2} (\delta_{da} f^a{}_{bc_{||}}  +\delta_{ba} f^a{}_{dc_{||}} )^2 \ ,
\eeq
where the square means here the contraction with two flat metrics of the two free indices $bd$. This contribution is thus negative. So a de Sitter solution requires
\beq
\mbox{de Sitter}:\quad \forall\ a_{||}\ ,\ \delta^{bd}\delta^{ce} f^{a_{||}}{}_{bc} f^{a_{||}}{}_{de} > 0  \Leftrightarrow \forall\ a_{||}\ ,\ \exists f^{a_{||}}{}_{bc} \neq 0  \ .\label{requirefabc}
\eeq
These new requirements help sharpening the possibilities left to find de Sitter solutions. For instance, in the failed attempt of appendix B of \cite{Andriot:2018ept}, \eqref{requirefabc} was verified but one had ${\cal R}_{11} = 0$ which was along the source: this could be a reason for the failure.

We finally consider transverse components ${}_{a_{\bot}b_{\bot}}$. The sign in that case is less definite due to the absence of $T_{ab}$ contribution. However, we need $0 > {\cal R}_6 = \delta^{ab} {\cal R}_{ab} = \delta^{a_{||}b_{||}} {\cal R}_{a_{||}b_{||}} + \delta^{a_{\bot}b_{\bot}} {\cal R}_{a_{\bot}b_{\bot}} $. Since the first contribution is positive, the second one $\delta^{a_{\bot}b_{\bot}} {\cal R}_{a_{\bot}b_{\bot}} $ should be negative. We deduce the requirement on components
\beq
\mbox{de Sitter}:\quad \exists\  a_{\bot}\ ,\  {\cal R}_{a_{\bot}a_{\bot}} < 0 \ .\label{reqRbot}
\eeq
Once again, the last term in the Ricci tensor expression \eqref{Ricci} is a square and the first two terms can be written as minus a square. We deduce that the latter should be non-zero, meaning
\beq
\mbox{de Sitter}:\quad \exists\  a_{\bot}\ ,\  \delta_{c(d} f^c{}_{b)a_{\bot}} \neq 0 \ .\label{reqfbot}
\eeq
The symmetry in free indices $bd$ forbids for instance algebras with fully antisymmetric structure constants, like semi-simple algebras in their standard basis, as already obtained in \cite{Andriot:2018ept}.

In view of these new constraints on structure constants and Ricci tensor components, we made a new attempt to find a solution with $p=5$. We considered a squashed solvmanifold with the algebra denoted ``s2.5'', ignoring for now the question of lattice existence (see e.g.~\cite{Grana:2006kf, Andriot:2015sia}). This allowed us to satisfy these various new sign requirements on algebra and Ricci tensor, contrary to the attempt in \cite{Andriot:2018ept}. A solution was nevertheless not found. From both attempts, it seemed in addition that only a subset of equations is enough to exclude de Sitter solutions: 4d, 10d traces of Einstein, dilaton e.o.m., and fluxes equations. We will however show in section \ref{sec:partcase} that this is not true.

It is difficult to get more useful information than obtained here, out of Einstein equation. The constraints or no-go theorems derived so far are based on linear combinations of equations to solve, and this method works as long as the quantities entering these equations are the same. So far, mostly scalar quantities were involved: squares of fluxes, curvatures or contractions of structure constants, and sources terms. The Einstein equation however involves tensor components. The same goes for fluxes e.o.m. and BI, expressed on forms and involving their components. This is the reason why it is in general difficult to obtain more useful information from these equations.\\

The only remaining equations are fluxes e.o.m.~and BI, given in section \ref{sec:sum}. For simplicity, we focus here on $p=6$ with $F_4=F_6=0$. The equations to consider are thus
\bea
& \d H=0\ , \ \d F_0=0 \ ,\ F_2 \w H =0 \ , \ \d F_2 - H \w F_0 = \frac{T_{10}}{p+1} \, {\rm vol}_{\bot} \ ,\\
& \d (*_6 H) = g_s^2 F_0 \w *_6 F_2 \ ,\ \d *_6 F_2 = 0 \ .
\eea
Several equations are automatic thanks to our ansatz: $F_0$ is constant so its BI is guaranteed; applying $\d$ on the $H$ e.o.m.~gives automatically the $F_2$ e.o.m., because $F_0$ has to be non-zero for a de Sitter solution. Finally, $F_2 \w H^{(0)} =0$ is automatic. As explained above, it is difficult to get useful information out these flux equations, especially scalar quantities. We nevertheless found here the following trick to get a scalar combination, incorporating equations we have not used so far, namely the $H$-flux e.o.m. and the $F_2$ BI along all directions.

We consider the top form $\d (F_2 \w *_6 H)$. With constant fluxes, this form is proportional to $f^a{}_{ab}$ which has to vanish due to compactness, so $\d (F_2 \w *_6 H) = 0$. We can develop it using both the $F_2$ BI and the $H$-flux e.o.m. This leads to the condition
\beq
\d (F_2 \w *_6 H) = 0 \quad \Rightarrow \quad |H|^2 + g_s^2 |F_2|^2 = - \frac{T_{10}}{p+1} \frac{(H^{(0)})_{\bot}}{F_0} \ , \label{fluxcond}
\eeq
where $H^{(0)} = (H^{(0)})_{\bot} {\rm vol}_{\bot} $. If $H^{(0)}=0$, then \eqref{fluxcond} gives $H=F_2=0$, which from the $F_2$ BI implies $T_{10}=0$, forbidden for a de Sitter solution. We deduce
\beq
\mbox{de Sitter}:\quad {\rm for}\ p=6\ {\rm with}\ F_4=F_6=0,\ \ H^{(0)} \neq 0 \ . \label{Hneq0par}
\eeq
In addition, since de Sitter solutions require $T_{10} > 0$, we deduce the sign
\beq
\mbox{de Sitter}:\quad {\rm for}\ p=6\ {\rm with}\ F_4=F_6=0,\ \ -(H^{(0)})_{\bot} F_0 > 0 \ . \label{signHF}
\eeq
Interestingly, this is the quantity that appears in the projected $F_2$ BI
\beq
(\d F_2)_{\bot} - (H^{(0)})_{\bot} F_0 = \frac{T_{10}}{p+1} \ . \label{Bibot}
\eeq
We use this expression back in \eqref{fluxcond} with $|H|^2  = |H^{(0)}|^2 +|H^{(2)}|^2 $, to obtain
\beq
|H^{(2)}|^2 + g_s^2 |F_2|^2 = - \frac{(H^{(0)})_{\bot}}{F_0}\, (\d F_2)_{\bot} \ , \label{dF2cond}
\eeq
from which we deduce the requirement
\beq
\mbox{de Sitter}:\quad {\rm for}\ p=6\ {\rm with}\ F_4=F_6=0,\ \ (\d F_2)_{\bot} \geq 0 \ . \label{dF2pos}
\eeq
This is interesting: it means that both contributions on the left-hand side of the $F_2$ BI \eqref{Bibot} have to be positive. In other words, they do not compete against each other, a result we will come back to in section \ref{sec:conj3}. Beyond these results, the condition \eqref{fluxcond} remains to difficult to use because it is not quadratic in the fluxes, and therefore hard to combine with a linear combination of the other equations. This illustrates again the difficulty in using flux equations. We nevertheless manage to go further in the next section.

\subsection{An interesting case}\label{sec:partcase}

There is a particular case of the previous study for which one can go further with the various equations
\beq
p=6,\ F_4=0,\ F_6=0,\ H^{(2)}=0 \ ,\label{simplecase}
\eeq
and to simplify notations we set $g_s=1$. In that case, the form $H=H^{(0)}=(H^{(0)})_{\bot} {\rm vol}_{\bot} $.\footnote{The simplicity of this setting might be extended to the other $p$ with $F_{k+2}=F_{k+4}=0$, $H^{(2)}=0$, and further split the transverse space into two orthogonal subspaces given by $H^{(0)}=(H^{(0)})_{\bot} {\rm vol}_{H} $ and $F_{k-2}=(F_{k-2})_{\bot} {\rm vol}_{k-2} $ , such that ${\rm vol}_{\bot}={\rm vol}_{H} \w {\rm vol}_{k-2}$, etc. One may then solve similarly the equations, and derive an analogous set of relations, considering e.g.~the top form $\d (F_k \w *_6(H\w F_{k-2}))$.} We denote the number $(H^{(0)})_{\bot} = h$, and $|H|^2=h^2$. Combining the dilaton e.o.m., the 4d and 10d Einstein traces, one obtains an expression for $T_{10}$ in terms of fluxes only. Combined with the projected BI \eqref{Bibot}, one deduces
\beq
(h+F_0) (2h-5F_0) = 3(|F_2|^2 - (\d F_2)_{\bot}) \ ,\ {\cal R}_4= \frac{2}{3} (F_0^2 - h^2) \ .
\eeq
Using condition \eqref{signHF} saying that $hF_0 <0$, we deduce the requirement
\beq
\mbox{de Sitter}:\quad {\rm for}\ p=6\ {\rm with}\ F_4=F_6=H^{(2)}=0,\ \ F_2 \neq 0 \ .
\eeq
We also deduce the need of $|F_2|^2 - (\d F_2)_{\bot} \neq 0$, noticed already in \cite{Andriot:2016xvq} in relation to Minkowski calibration of sources. Combining with \eqref{dF2cond}, we obtain
\beq
|F_2|^2 = - \frac{h}{F_0}\, (\d F_2)_{\bot} = \frac{1}{3} (2h^2-5hF_0) \ ,\ \frac{T_{10}}{p+1}=  \frac{5}{3} (-hF_0 + F_0^2) \ ,\ {\cal R}_6 = -\frac{1}{6} (9 F_0^2 - 7h^2 - 5 hF_0) \ .\label{expr}
\eeq
The trace of the Einstein equation along internal parallel directions now gives
\beq
{\cal R}_{||} + {\cal R}_{||}^{\bot} + \frac{1}{2} |f^{{}_{||}}{}_{{}_{\bot} {}_{\bot}}|^2 = \frac{1}{6} (3F_0^2 - 5hF_0 - h^2) \ ,
\eeq
and we recall the following expression on group manifolds
\beq
{\cal R}_6 = {\cal R}_{||} + {\cal R}_{||}^{\bot} - \frac{1}{2} |f^{{}_{||}}{}_{{}_{\bot} {}_{\bot}}|^2  - \delta^{cd}   f^{b_{\bot}}{}_{a_{||}  c_{\bot}} f^{a_{||}}{}_{ b_{\bot} d_{\bot}} =  {\cal R}_{||} + {\cal R}_{||}^{\bot} + ( \lambda - \frac{1}{2}) |f^{{}_{||}}{}_{{}_{\bot} {}_{\bot}}|^2   \ , \label{R6group}
\eeq
where $0 < \lambda < 1$ for de Sitter solutions \cite{Andriot:2018ept}. We deduce
\beq
( \lambda - 1) |f^{{}_{||}}{}_{{}_{\bot} {}_{\bot}}|^2 = \frac{1}{3} (- 6 F_0^2 + 5 hF_0 + 4h^2) \ , \ 2({\cal R}_{||} + {\cal R}_{||}^{\bot}) = - \lambda |f^{{}_{||}}{}_{{}_{\bot} {}_{\bot}}|^2 + h^2 - F_0^2 \ . \label{expreR}
\eeq
We recover the requirement ${\cal R}_{||} + {\cal R}_{||}^{\bot} < 0$ for a de Sitter solution in the case $H^{(2)}=0 $ \cite{Andriot:2016xvq}.

Remarkably, we have managed so far to solve all equations considered in terms of $h, F_0$, even though fixing a geometry could eventually lead to constraints with the curvature terms. The only equations left to solve are the $H$-flux e.o.m. (not fully used so far), the $F_2$ BI projected along non-transverse directions, and the Einstein equation (we solved traces of it).\footnote{The $H$ BI is automatically satisfied as long as we use standard algebra basis, where $f^a{}_{ab} =0$ without sum on $a$ (see \cite{Andriot:2010ju}). Indeed, $\d H$ is here proportional to  $f^{a_{\bot}}{}_{a_{\bot} b_{||}}$.} We first get here from the $H$-flux e.o.m.
\bea
F_2&=-\frac{h}{F_0} *_6 \d {\rm vol}_{||} = -\frac{h}{F_0} *_6 \left( -\frac{1}{2}f^{a_{||}}{}_{b_{\bot} c_{\bot}}\ \iota_{a_{||}}\! {\rm vol}_{||} \w e^{b_{\bot}}\w e^{c_{\bot}}  \right) \\
&= \frac{h}{F_0} \frac{1}{2}f^{a_{||}}{}_{b_{\bot} c_{\bot}}\ \epsilon^{b_{\bot} c_{\bot}}{}_{d_{\bot}} \delta_{a_{||}e_{||}}\ e^{e_{||}} \w e^{d_{\bot}} \ ,\nn\\
\Rightarrow |F_2|^2 &=  \frac{h^2}{F_0^2}\ |f^{{}_{||}}{}_{{}_{\bot} {}_{\bot}}|^2 \ ,\\
F_{2\ a_{||}c}F_{2\ b_{||}}^{\ \ \ \ c}  &= \frac{h^2}{F_0^2}\ \frac{1}{2} \delta^{gd}\delta^{ce} f^{a_{||}}{}_{gc} f^{b_{||}}{}_{de} \ . \label{F2compo}
\eea
We deduce with $|F_2|^2$ the following relations
\beq
|f^{{}_{||}}{}_{{}_{\bot} {}_{\bot}}|^2  = \frac{1}{3}\, \frac{F_0^2}{h^2}\ (2h^2-5hF_0) \ , \  1-\lambda = \frac{h^2}{F_0^2}\ \frac{6 F_0^2 - 5 hF_0 - 4h^2}{2h^2-5hF_0} \ . \label{f2lam}
\eeq
The second and last flux equation to consider is the $F_2$ BI along non-transverse directions. In absence of $H^{(2)}$ (which otherwise gets defined from it), this identity gives an important constraint among structure constants
\bea
& \d F_2|_{\neq \bot} =  \frac{h}{F_0} \frac{1}{2}f^{a_{||}}{}_{b_{\bot} c_{\bot}}\ \epsilon^{b_{\bot} c_{\bot}}{}_{d_{\bot}} \delta_{a_{||}e_{||}}\ \left( - \frac{1}{2} f^{e_{||}}{}_{f_{||} g_{||}} \delta^{d_{\bot}}_{h_{\bot}} + \delta^{e_{||}}_{f_{||}} f^{d_{\bot}}{}_{g_{||} h_{\bot} } \right)  e^{f_{||}} \w e^{g_{||}} \w e^{h_{\bot}} = 0 \nn\\
& \Leftrightarrow \ \frac{1}{2}f^{a_{||}}{}_{b_{\bot} c_{\bot}}\ \left( - \delta_{a_{||}e_{||}} f^{e_{||}}{}_{f_{||} g_{||}} \epsilon^{b_{\bot} c_{\bot}}{}_{h_{\bot}} + 2 \epsilon^{b_{\bot} c_{\bot}}{}_{d_{\bot}} \delta_{a_{||}[f_{||}}  f^{d_{\bot}}{}_{g_{||}] h_{\bot} } \right) = 0 \ .\label{dF2nonbot}
\eea

At this stage, the problem boils down to having $h,F_0$ such that $F_0^2> h^2$, $hF_0 <0$, and having an algebra of a compact group manifold solving \eqref{f2lam}, the expression for ${\cal R}_{||} + {\cal R}_{||}^{\bot}$ in \eqref{expreR}, and the constraint \eqref{dF2nonbot}. In that case, all flux equations are solved, as well as usually considered equations, namely the dilaton e.o.m., the 4d and 10d Einstein traces, the trace along internal parallel directions. The only equations left are then internal Einstein equations, with traces already taken into account. Let us give the following explicit example
\bea
& \mbox{$O_6$ along directions } 123, \ f^1{}_{23}=-f_1,\ f^3{}_{45}=-f_3,\ f^4{}_{35}=-f_4 \ ,\\
\Rightarrow\ & |f^{{}_{||}}{}_{{}_{\bot} {}_{\bot}}|^2= f_3^2 ,\ \lambda=-\frac{f_4}{f_3},\ {\cal R}_{||} = -\frac{1}{2} f_1^2,\ {\cal R}_{||}^{\bot}=-\frac{1}{2} f_4^2 \ .
\eea
This solvmanifold is very likely to be compact: few constraints have to be imposed on the structure constants for it to admit a lattice, but given our conventions make the radii enter the structure constants, there is probably enough freedom to satisfy the (lattice) quantization conditions.\footnote{For instance, for the set of values \eqref{value}, the $\sqrt{3}$ of $f_1$ can be accommodated with the radius along $1$, that appears in no other structure constant or flux.} It is easy to verify that \eqref{dF2nonbot} is satisfied, in particular $F_2=-f_3 h/F_0\, e^3 \w e^6$. We then pick the following values which allow to satisfy the above, in particular \eqref{f2lam} and \eqref{expreR} with an appropriate $\lambda$ value
\beq
F_0=2,\ h=-1,\ f_3=4,\ f_4=-\frac{3}{2},\ f_1=\frac{3\sqrt{3}}{2} \ .\label{value}
\eeq
This shows that the information of each internal Einstein equation (and not only their trace) is required to prove conjecture \ref{conj1}. And indeed here, it is clear that they will not admit a de Sitter solution, because $f^2{}_{bc} =0$, violating \eqref{requirefabc}.

The expression \eqref{F2compo} of $F_2$ components is useful in view of solving the internal Einstein equations. Note also that this combination of structure constants appears in the Ricci tensor components. The problem then boils down to a set of quadratic equations on structure constants and $h,F_0$. In a concrete example, using condition \eqref{dF2nonbot} and subtracting one Einstein equation from another one, both parallel to the source, we could conclude on the absence of solutions. We refrain from trying such a method generically here. The question of proving conjecture \ref{conj1} therefore remains open, even in the simple case \eqref{simplecase}.

\section{Conjecture 2: solutions with intersecting sources are unstable}

Intersecting $O_p/D_p$ sources are grouped in sets labeled by $I=1 \dots N$ of parallel sources. In the equations, the only difference between the case of parallel ($N=1$) and intersecting ($N>1$) sources appears in the source contributions, i.e.~in $T_{MN}$ and its trace $T_{10}= \sum_{I=1}^N T_{10}^I$. We refer to sections \ref{sec:ansatz}, \ref{sec:sum} and appendix \ref{ap:eq} for relevant definitions. The overlap between different sets $I$ and $J$ plays an important role, captured by the numbers $\delta_{a_{||_I}}^{a_{||_J}}$ of common directions between the two sets. We restrict in section \ref{sec:conj2attempt} to the case of homogeneous overlap \cite{Andriot:2017jhf} where all these numbers are identical
\beq
\mbox{Homogeneous overlap:} \ \forall I,\, J\neq I,\ \delta_{a_{||_I}}^{a_{||_J}} = N_o\ \mbox{independent of}\ I,\, J\ .\label{homoov}
\eeq
This is the situation of all examples of de Sitter solutions with intersecting $O_6/D_6$, with $N_o=1$. The difference in equations due to source contributions is responsible for having found de Sitter solutions with intersecting sources and not with parallel ones: as recalled in section \ref{sec:existinter}, existence constraints coming from the trace of the Einstein equation along internal parallel directions are relaxed for intersecting sources \cite{Andriot:2017jhf}. An intuition is that the new source contributions, while allowing for solutions, lead at the same time to an instability or systematic tachyon, as defined in section \ref{sec:stab}. If this holds, a proof of conjecture \ref{conj1} could provide a proof of conjecture \ref{conj2}. Following this idea, we developed similar tools and strategy for both problems: we present them in sections \ref{sec:pot} and \ref{sec:conj2attempt}. They allow us to find further existence constraints in section \ref{sec:existinterfurther}.

\subsection{Existence conditions}\label{sec:existinter}

Before tackling the stability of classical de Sitter solutions with intersecting sources, we first review known existence conditions, and find new ones. The dilaton e.o.m., the 4d and 10d traces of the Einstein equation are formally the same as those for parallel sources; they correspond in 4d language to the 4d Einstein equation and $\del_{\rho} V = \del_{\tau} V = 0$, given in \eqref{eqR4}, \eqref{eqdeltauV}, \eqref{eqdelrhoV}. So we obtain from them the same constraints as in the parallel case \cite{Wrase:2010ew, Shiu:2011zt, Andriot:2016xvq}. Considering additionally the $F_k$ BI \eqref{BI2} brings further restrictions, excluding $p=3$, and the $F_{k-4}$ BI leads to $F_0=0$ for $p=4$ as for parallel sources \cite{Andriot:2018ept}, excluding further cases.

The difference with the case of parallel sources then appears when considering the trace of Einstein equation along internal parallel directions. The extra source contributions, obtained in 10d in \cite{Andriot:2017jhf} and in 4d in \eqref{delVsigma}, relax the constraint on $f^{||}{}_{\bot\bot}$: these structure constants do not have to be non-zero anymore. The constraint on the NS combination of curvatures and $H$-flux components \cite{Andriot:2016xvq}, denoted ``combi'', still got generalized to the intersecting case \cite{Andriot:2017jhf}, at least for a homogeneous overlap.\footnote{Another constraint obtained in \cite{Andriot:2017jhf} from this NS combination on group manifolds with constant fluxes is the need, for $p=6$, of some sources overlap: a de Sitter solution with $N_o=0$, implying $N=2$, is excluded.} This leaves the possibilities and necessary ingredients summarized in Table \ref{tab:inters}. The difference in the Einstein along parallel directions, with a relaxed requirement on $f^{||}{}_{\bot\bot}$, make it unlikely to get existence constraints in terms of the parameter $\lambda$ \eqref{lambda}, discussed in section \ref{sec:c}. So we do not get a further requirement of having some non-zero $f^{\bot}{}_{||\bot}$ \cite{Andriot:2018ept}. Similarly, the new constraints derived in section \ref{sec:othereqpar} for parallel sources, namely \eqref{reqRpar}, \eqref{requirefabc}, \eqref{reqRbot} and \eqref{reqfbot}, cannot be obtained here. There is nevertheless a very special case of intersecting sources, where the requirement on $f^{||}{}_{\bot\bot}$ does not get relaxed, so one can then generalize the constraints on $\lambda$: we study this in section \ref{sec:existinterfurther}, after having defined required tools.

\begin{table}[h]
\begin{center}
\begin{tabular}{|c|c|c|}
    \hline
     & \multicolumn{2}{|c|}{A de Sitter solution requires $T_{10}>0$ and}\\
    \hline
    $p=\dots$ & ${\cal R}_6 \geq 0$  & ${\cal R}_6 <0$ \\
    \hline
    \hline
    3 &  &  \\
    \hline
    4 &  & $\ \ F_2$, $ \mbox{combi}\ \ $ \\
    \hline
    5 &  & $F_1$, $ \mbox{combi}$ \\
    \hline
    6 &  & $F_0$, $ \mbox{combi}$ \\
    \hline
    7 &  &  \\
    \hline
    8 &  &  \\
    \hline
    9 &  &  \\
    \hline
\end{tabular}\caption{Necessary ingredients for classical de Sitter solutions with intersecting $O_p/D_p$ sources, prior to this work. An empty box means a no-go theorem.}\label{tab:inters}
\end{center}
\end{table}

The new constraints obtained in section \ref{sec:othereqpar} from the flux equations can on the contrary be generalized in any case to intersecting sources. For $p=6$ with $F_4=F_6=0$, using $F_2$ BI \eqref{BI2}, we obtain as in \eqref{fluxcond}
\beq
\d (F_2 \w *_6 H) = 0 \quad \Rightarrow \quad |H|^2 + g_s^2 |F_2|^2 = - \sum_I \frac{T_{10}^I}{p+1} \frac{(H^{(0)_I})_{\bot_I}}{F_0} \ . \label{fluxcondinter}
\eeq
Consequences are however weaker, due to the sum on $I$. For instance, one requirement is that $T_{10} > 0$, but each $T_{10}^I$ does not need to be. We deduce
\bea
\mbox{de Sitter}: & \quad {\rm for}\ p=6\ {\rm with}\ F_4=F_6=0,\ \ \exists\ I\ {\rm s.t.}\ T_{10}^I\, (H^{(0)_I})_{\bot_I} \neq 0 \ , \label{Hneq0inter}\\
\mbox{de Sitter}: & \quad {\rm for}\ p=6\ {\rm with}\ F_4=F_6=0,\ \ \exists\ I\ {\rm s.t.}\ - T_{10}^I\, (H^{(0)_I})_{\bot_I}\, F_0 > 0 \ .\label{signHFinter}
\eea
Furthermore, the projected BI gives
\beq
\forall J,\ (\d F_2)_{{\bot}_J} - (H^{(0)_J})_{{\bot}_J}\, F_0 = \frac{T_{10}^J}{p+1} \ . \label{BibotJ}
\eeq
Using this back in \eqref{fluxcondinter}, we obtain for any given $I$
\beq
|H^{(2)_I}|^2 - \sum_{J\neq I} |H^{(0)_J}|^2  + g_s^2 |F_2|^2 =  - \sum_J (\d F_2)_{{\bot}_J}\, \frac{(H^{(0)_J})_{\bot_J}}{F_0} \ .\label{dF2coninter}
\eeq
The left hand-side is positive. Indeed, $\forall I,J$, $H= H^{(0)_I}+ H^{(2)_I} = H^{(0)_J}+ H^{(2)_J}$. Since $p=6$, $\forall K$, $H^{(0)_K}$ is proportional to ${\rm vol}_{{\bot}_K}$ which is by definition along different directions for each $K$, so each $H^{(0)_J}$ is necessarily a piece of $H^{(2)_I}$. Since each $H^{(0)_J}$ is along different directions, one deduces $|H^{(2)_I}|^2 - \sum_{J\neq I} |H^{(0)_J}|^2 \geq 0$. The right hand-side of \eqref{dF2coninter} is thus positive, and we deduce the generalization of \eqref{dF2pos}
\beq
\mbox{de Sitter}:\quad {\rm for}\ p=6\ {\rm with}\ F_4=F_6=0,\ \ \exists\ I\ {\rm s.t.}\ -(\d F_2)_{{\bot}_I}\, (H^{(0)_I})_{\bot_I}\, F_0 \geq 0 \ . \label{dF2posinter}
\eeq
So there is at least one set $I$ for which both contributions on the left-hand side of the projected $F_2$ BI \eqref{BibotJ} have the same sign, i.e.~are not competing against each other. These new requirements for the existence of classical de Sitter solutions with intersecting sources are interesting, but it remains hard to use them further, and generalize the study of section \ref{sec:partcase}.

\subsection{Strategy and tools to prove instability}\label{sec:pot}

The strategy to prove conjecture \ref{conj2} follows an idea proposed and checked on some examples in \cite{Danielsson:2012et}, and further checked in \cite{Junghans:2016uvg}. The idea is that studying the stability of the scalar fields $\rho, \tau, \sigma$ (extended to $\sigma_I$ for intersecting sources) should be enough to find a tachyon. In other words, a combination of these fields should always be at a maximum of the potential $V$, providing a systematic tachyon thus proving conjecture \ref{conj2}. To that end, one should consider the block of the Hessian of the potential corresponding to this subset of fields, and prove it always admits a negative eigenvalue. This seems simpler than dealing with the full 4d supergravity scalar potential (see section \ref{sec:stab}). The problem of finding a systematic negative eigenvalue in the Hessian block is further simplified using Sylvester criterion (see e.g.~\cite{Shiu:2011zt}): if one of the diagonal entries, e.g.~$\del_{\rho}^2 V$, is always negative, this is sufficient to conclude on a negative eigenvalue. The strategy \cite{Andriot:2018ept} then consists in building a linear combination of terms that have to be positive or zero on a de Sitter solution (${\cal R}_4$, $\del_{\varphi} V$, $|\dots|^2$, etc.) and of $\del^2_{\varphi} V$ times positive coefficients: see e.g.~\eqref{galLC3}. Proving that this linear combination is always negative, one would deduce that there is always a negative diagonal entry, and conclude. This idea of the linear combination is in line with previous no-go theorems, discussed in section \ref{sec:c}.\\

The first step is to determine the scalar potential $V(\rho, \tau, \sigma_I)$. This was done in \cite{Andriot:2018ept} for parallel sources, i.e.~with one $\sigma$, and we follow here closely this derivation to generalize it to intersecting sources. As explained in section \ref{sec:ansatz}, each set $I=1 \dots N$ of parallel sources defines internal directions $a_{||_I}$ parallel to these sources, and transverse ones $a_{\bot_I}$, in the orthonormal (flat) basis with metric $\delta_{ab}$. For any set $I$, we then write the 6d internal metric as
\beq
\d s_6^2 = \delta_{ab} e^a e^b = \d s_{||_I}^2 + \d s_{\bot_I}^2 \ ,\quad  \d s_{||_I}^2 \equiv e^{a_{||_I}}{}_{m} e^{b_{||_I}}{}_{n} \delta_{ab} \d y^m \d y^n \ ,\quad \d s_{\bot}^2 \equiv e^{a_{\bot_I}}{}_{m} e^{b_{\bot_I}}{}_{n} \delta_{ab} \d y^m \d y^n \ .\nn
\eeq
Each set $I$ admits two sets of one-forms $\{e^{a_{||_I}}\}, \{e^{a_{\bot_I}}\}$, where the assumptions on those, detailed in section \ref{sec:ansatz}, hold for group manifolds. The 4d scalar fields $\rho, \tau, \sigma_{I=1\dots N} >0$ are obtained from fluctuations of 10d fields around a background labeled with a ${}^0$. They are defined as follows for each set $I$
\beq
\hspace{-0.25cm} \d s_6^2 = \rho \left( \sigma_I^A (\d s_{||_I}^2)^0 + \sigma_I^B (\d s_{\bot_I}^2)^0 \right) \ , \ e^{a_{||_I}}{}_{m} = \sqrt{\rho \sigma_I^A}\, (e^{a_{||_I}}{}_{m})^0 \ ,\ e^{a_{\bot_I}}{}_{m} = \sqrt{\rho \sigma_I^B}\, (e^{a_{\bot_I}}{}_{m})^0 \label{fluct}
\eeq
with $A= p-9$, $B=p-3$. The dilaton fluctuation is defined by $\phi = \phi^0 + \delta \phi$ with $\tau = e^{- \delta \phi} \rho^{\frac{3}{2}}$, where $e^{\phi^0}=g_s$ is a constant. Background values are recovered with
\beq
\mbox{Background value:}\quad \rho= \tau = \sigma_{I=1\dots N} =1 \ . \label{bckgdN}
\eeq
In \eqref{fluct}, all $\sigma_{I=1\dots N}$ should actually appear at the same time; we froze to their background value all but one to simplify the definition. The 6d metric determinant fluctuates as
\beq
|g_6| \rightarrow |g_6^0|\ \rho^6\ \Pi_{I=1}^N\ \sigma_I^{A (p-3) + B (9-p)} = |g_6^0|\ \rho^6 \ , \label{metricdet}
\eeq
where the $\sigma_I$ dependence drops out thanks to the values picked for $A$ and $B$. This justifies the definition of $\tau$, and allows to define the 4d Planck mass and the relation to the 4d Einstein frame as in the parallel sources case.

From there, determining the scalar potential amounts to fluctuate the curvature, fluxes and sources. We refer to appendix \ref{ap:pot} for details of this derivation. As expected, only the source contributions are significantly different from that of the parallel case, and they depend on their possible overlap. The resulting potential is given as follows
\begin{empheq}[innerbox=\fbox, left=\!\!\!\!\!\!\!\!\!\!\!\!\!\!\!\!\!\!]{align}
\tV = \frac{V}{M_4^2} = & - \tau^{-2} \bigg( \rho^{-1} {\cal R}_6(\sigma_K) -\frac{1}{2} \rho^{-3} \sum_n \sigma_I^{-An-B(3-n)} |H^{(n)_I}|^2(\sigma_J) \bigg) \label{pot}\\
& - g_s \tau^{-3} \rho^{\frac{p-6}{2}} \sum_I \sigma_I^{A\frac{p-3}{2}} \Pi_{J\neq I}\, \sigma_J^{\frac{(A-B)\delta_{a_{||_I}}^{a_{||_J}} + B (p-3)}{2}}   \frac{T_{10}^I}{p+1} \nn\\
& \hspace{-0.7in} +\frac{1}{2} g_s^2 \bigg( \tau^{-4} \sum_{q=0}^{4} \rho^{3-q} \sum_n  \sigma_I^{-An-B(q-n)} |F_q^{(n)_I}|^2(\sigma_J)  - \tau^{4} \rho^3 |F_6|^2 \nn\\
& \hspace{-0.7in} \phantom{+\frac{1}{2} g_s^2  \bigg(}\!\! + \frac{1}{2}  \sum_n   (\tau^{-4} \rho^{-2} \sigma_I^{-An-B(5-n)} |F_5^{(n)_I}|^2(\sigma_J) - \tau^{4} \rho^2 \sigma_I^{-An-B(1-n)} |(*_6 F_5)^{(n)_I}|^2(\sigma_J) ) \bigg) \nn
\end{empheq}
where only even/odd RR fluxes should be considered in IIA/IIB, and we use simplified notations absorbing the 6d integral and dropping the background label
\beq
\frac{\int \d y^6 \sqrt{|g_6^0|} |H^{(n)_I 0}|^2 }{\int \d y^6 \sqrt{|g_6^0|}} \rightarrow |H^{(n)_I 0}|^2 \rightarrow |H^{(n)_I}|^2 \ .
\eeq
This potential is derived using only part of the ansatz of section \ref{sec:ansatz}, namely the absence of warp factor and a constant dilaton; we did not restrict to group manifolds with constant fluxes, and the general curvature dependence ${\cal R}_6(\sigma_K)$ is given in appendix \ref{ap:pot}. The important novelty in the source terms is not the extra dependence on $\sigma_J$: this happens naturally in all terms of the potential. Rather, it is the presence of $\sigma_I$ in all source terms, and not only the one with $T_{10}^I$. This generates new source contributions with respect to the case of parallel sources.\\

The next step is to derive 4d equations. The potential, and its first and second derivatives with respect to $\rho, \tau$ are the same as in the parallel case when setting all $\sigma_I=1$. This is a relevant value since we will consider derivatives of the potential at an extremum, so with the background value \eqref{bckgdN}. So the 4d Einstein equation and the two first derivatives are the same as in the parallel case, and we recall them for completeness
\bea
{\cal R}_4= &\ -  2 {\cal R}_6 + |H|^2  - 2 g_s \frac{T_{10}}{p+1}  + g_s^2 \sum_{q=0}^{6} |F_q|^2  \label{eqR4} \\
 0 = \tau \del_\tau \tV|_0 = &\ 2  {\cal R}_6 -  |H|^2 +3 g_s \frac{T_{10}}{p+1} -2 g_s^2  \sum_{q=0}^{6} |F_q|^2 \label{eqdeltauV}\\
 0 = \rho \del_\rho \tV|_0 = &\  {\cal R}_6 - \frac{3}{2} |H|^2  - g_s \frac{p-6}{2} \frac{T_{10}}{p+1}  +\frac{1}{2} g_s^2 \sum_{q=0}^{6} (3-q) |F_q|^2  \label{eqdelrhoV} \ .
\eea
These equations are equivalent to the (6d integral of) the 4d and 10d traces of 10d Einstein equation and the 10d dilaton e.o.m. \cite{Andriot:2018ept}, which are the same in the parallel or intersecting case (see section \ref{sec:existinter}). The same constraints can then be deduced, such as
\beq
{\cal R}_4= g_s \frac{T_{10}}{p+1} - g_s^2 \sum_{q=0}^{6} |F_q|^2 \label{R4T10F} \ ,
\eeq
which implies the requirement $T_{10}>0$, i.e.~the need for $O_p$ in at least one set $I$ \cite{Andriot:2017jhf}. Finally, we should obtain the derivatives with respect to $\sigma_I$: we do so, and comment on them, in appendix \ref{ap:pot}. Following the strategy presented at the beginning of this section, we now consider a linear combination of all these equations.

\subsection{Attempt in a standard case}\label{sec:conj2attempt}

We have developed tools and presented a strategy to prove conjecture \ref{conj2}, and we now apply those to a standard case where most classical de Sitter solutions were found \cite{Danielsson:2011au}, all being unstable. The setting matches the ansatz described in section \ref{sec:ansatz}: we consider group manifolds with constant fluxes, for $p=6$. On a group manifold with an orientifold along a set $I$, ${\cal R}_6$ simplifies to (see \eqref{R6group})
\beq
{\cal R}_6 = {\cal R}_{||_I} + {\cal R}_{||_I}^{\bot_I} - \frac{1}{2} |f^{{}_{||_I}}{}_{{}_{\bot_I} {}_{\bot_I}}|^2  - \delta^{cd}   f^{b_{\bot_I}}{}_{a_{||_I}  c_{\bot_I}} f^{a_{||_I}}{}_{ b_{\bot_I} d_{\bot_I}}  \ . \label{R6gp}
\eeq
In addition, for $p=6$, one gets for the RR fields $(A-B)n + qB = 0$, i.e.~no contribution of those in the $\del_{\sigma_I}$. Finally, intersecting sources are now in the case of homogenous overlap \eqref{homoov}, with $N_o = 1$. The derivatives simplify to
\bea
\sigma_I \del_{\sigma_I} \tV|_0 = & \ -3 \left( {\cal R}_{||_I} +  {\cal R}_{||_I}^{\bot_I} \right)  - \frac{9}{2} |f^{{}_{||_I}}{}_{{}_{\bot_I} {}_{\bot_I}}|^2 - 3\, \delta^{cd}   f^{b_{\bot_I}}{}_{a_{||_I}  c_{\bot_I}} f^{a_{||_I}}{}_{ b_{\bot_I} d_{\bot_I}}   \label{delVsigma62}\\
& - \frac{1}{2}  ( 9 |H^{(0)_I}|^2- 3 |H^{(2)_I}|^2) + g_s \left( 9 \frac{T_{10}^I}{p+1} + 3 \sum_{J\neq I} \frac{T_{10}^J}{p+1}   - \frac{9}{2} \frac{T_{10}}{p+1} \right) \ , \nn\\
\sigma_I^2 \del^2_{\sigma_I} \tV|_0 = & \ -\sigma_I \del_{\sigma_I} \tV|_0  - 9 \left( {\cal R}_{||_I} +  {\cal R}_{||_I}^{\bot_I} \right)  +\frac{9^2}{2}  |f^{{}_{||_I}}{}_{{}_{\bot_I} {}_{\bot_I}}|^2 + 9 \, \delta^{cd}   f^{b_{\bot_I}}{}_{a_{||_I}  c_{\bot_I}} f^{a_{||_I}}{}_{ b_{\bot_I} d_{\bot_I}} \nn \\
& + \frac{1}{2} ( 9^2 |H^{(0)_I}|^2 + 9 |H^{(2)_I}|^2 )  - \frac{g_s}{4} \bigg( 9^2  \frac{T_{10}^I}{p+1} + 9 \sum_{J\neq I}  \frac{T_{10}^J}{p+1}  \bigg) \ . \label{deldelVsigma62}
\eea
The $F_{k=2}$ BI \eqref{BI2}, projected on one ${\rm vol}_{\bot_I}$ as in \eqref{BibotJ}, can also be rewritten as follows \cite{Andriot:2017jhf} (using that $F_{k}^{(0)_I}=0$)
\bea
g_s \frac{T_{10}^I}{p+1} = & - \frac{1}{2} \left|*_{\bot_I}H^{(0)_I} +  g_s F_0 \right|^2  - \frac{1}{2} \sum_{a_{||_I}} \left| *_{\bot_I}( \d e^{a_{||_I}})|_{\bot_I} - g_s\, \iota_{a_{||_I}} F_2^{(1)_I} \right|^2 \label{BI} \\
& + \frac{1}{2}|H^{(0)_I}|^2 + \frac{1}{2} g_s^2 |F_0|^2 + \frac{1}{2} g_s^2 |F_{2}^{(1)_I}|^2 + \frac{1}{2} |f^{{}_{||_I}}{}_{{}_{\bot_I} {}_{\bot_I}}|^2 \ . \nn
\eea
We now build a general linear combination as in \cite{Andriot:2018ept}, following the strategy presented in section \ref{sec:pot}. With real coefficients $a, b_{\tau}, b_{\rho}, b_{\sigma}, c_{\tau}, c_{\rho}, c_{\sigma}, d$, we consider
\bea
& a {\cal R}_4 + b_{\tau} \ \tau \del_{\tau} \tV|_0  + b_{\rho} \ \rho \del_{\rho} \tV|_0  + b_{\sigma} \ \sigma_I \del_{\sigma_I} \tV|_0 \label{galLC3}\\
& + c_{\tau} \ \tau^2 \del^2_{\tau} \tV|_0 + c_{\rho} \ \rho^2 \del^2_{\rho} \tV|_0 + c_{\sigma} \ \left(  \sigma_I^2 \del^2_{\sigma_I} \tV|_0 + \sigma_I \del_{\sigma_I} \tV|_0 \right)\nn\\
& +\frac{d}{2} \left|*_{\bot_I}H^{(0)_I} +  g_s F_0 \right|^2  + \frac{d}{2} \sum_{a_{||_I}} \left| *_{\bot_I}( \d e^{a_{||_I}})|_{\bot_I} -  g_s\, \iota_{a_{||_I}} F_2^{(1)_I} \right|^2 \nn\\
= & - ({\cal R}_{||_I} + {\cal R}_{||_I}^{\bot_I})\ (2a-2b_{\tau} -b_{\rho} -A b_{\sigma} + 6 c_{\tau} + 2 c_{\rho} + A^2 c_{\sigma}) \nn\\
& + \delta^{cd}   f^{b_{\bot_I}}{}_{a_{||_I}  c_{\bot_I}} f^{a_{||_I}}{}_{ b_{\bot_I} d_{\bot_I}}\ (2a-2b_{\tau} -b_{\rho} -B b_{\sigma} + 6 c_{\tau} + 2 c_{\rho} + B^2 c_{\sigma}) \nn\\
& + \frac{1}{2} |f^{{}_{||_I}}{}_{{}_{\bot_I} {}_{\bot_I}}|^2\  (2a-2b_{\tau} -b_{\rho} +(A-2B) b_{\sigma} + 6 c_{\tau} + 2 c_{\rho} + (B^2 +(B-A)(3B-A) ) c_{\sigma} + d ) \nn\\
& +\frac{1}{2} |H^{(0)_I}|^2 \ (2a-2b_{\tau} -3b_{\rho} -3B b_{\sigma} + 6 c_{\tau} + 12 c_{\rho} + 9B^2 c_{\sigma} +d) \nn\\
& +\frac{1}{2} |H^{(2)_I}|^2 \ (2a-2b_{\tau} -3b_{\rho} -(2A+B) b_{\sigma} + 6 c_{\tau} + 12 c_{\rho} + (2A+B)^2 c_{\sigma}) \nn\\
& + g_s \frac{T_{10}^I}{p+1} \frac{1}{4}\ (-8a +12b_{\tau} +18 b_{\sigma} -48 c_{\tau} -9^2 c_{\sigma}-4d) \nn\\
& + g_s \sum_{J\neq I} \frac{T_{10}^J}{p+1} \frac{1}{4}\ (-8a +12b_{\tau} -6 b_{\sigma} -48 c_{\tau} -9 c_{\sigma}) \nn\\
& + g_s^2 \sum_{q=0}^4 \frac{1}{2} |F_q|^2\ (2a-4b_{\tau} + (3-q) b_{\rho}  + 20 c_{\tau} + (3-q)(2-q) c_{\rho})  + g_s^2 \frac{d}{2} ( |F_0|^2 + |F_{2}^{(1)_I}|^2 ) \nn\\
& + g_s^2 \frac{1}{2} |F_6|^2\ (2a-4b_{\tau}  -3 b_{\rho}  -12 c_{\tau} -6 c_{\rho}) \nn
\eea
The only difference with respect to the parallel case are the $T_{10}^J$ terms. As explained in section \ref{sec:pot}, the left-hand side is strictly positive on a (meta)stable de Sitter solution, when restricting to $a>0$ and $ c_{\tau}, c_{\rho}, c_{\sigma}, d \geq 0$. Therefore, if we can find coefficients such that the right-hand side is negative or zero, we get a no-go theorem. We know the sign of most entries on the right-hand side, e.g.~$|H^{(0)_I}|^2 \geq 0$, so we should find coefficients $a,b,c,d$ such that the associated combination, $2a-2b_{\tau} -3b_{\rho} -3B b_{\sigma} + 6 c_{\tau} + 12 c_{\rho} + 9B^2 c_{\sigma} +d$, is negative or zero. This would build the appropriate linear combination to get a no-go theorem, and prove conjecture \ref{conj2} in this standard case. This strategy was successful for parallel sources, to get existence constraints: the corresponding linear combinations \eqref{l<0}, \eqref{l>1}, were discussed in section \ref{sec:c}. With the above general linear combination however, we do not reach any conclusive result regarding stability. We will nevertheless deduce further existence constraints in section \ref{sec:existinterfurther}.

We then consider two different combinations, first
\bea
& a N {\cal R}_4 + b_{\tau} N \ \tau \del_{\tau} \tV|_0  + b_{\rho} N \ \rho \del_{\rho} \tV|_0  + b_{\sigma} \ \sum_I \sigma_I \del_{\sigma_I} \tV|_0 \label{galLC}\\
& + c_{\tau} N \ \tau^2 \del^2_{\tau} \tV|_0 + c_{\rho} N \ \rho^2 \del^2_{\rho} \tV|_0 + c_{\sigma} \sum_I\ \left(  \sigma_I^2 \del^2_{\sigma_I} \tV|_0 + \sigma_I \del_{\sigma_I} \tV|_0 \right)\nn\\
& +\frac{d}{2} \sum_I \left|*_{\bot_I}H^{(0)_I} + \varepsilon_p g_s F_{k-2}^{(0)_I} \right|^2  + \frac{d}{2} \sum_I \sum_{a_{||_I}} \left| *_{\bot_I}( \d e^{a_{||_I}})|_{\bot_I} - \varepsilon_p g_s\, \iota_{a_{||_I}} F_k^{(1)_I} \right|^2 \ ,\nn
\eea
where essentially, we summed over all $I$ or equivalently multiplied by $N$. This allows to reconstruct $T_{10}$ using e.g.~that $\sum_I \sum_{J\neq I} T_{10}^J = (N-1) T_{10}$. We then obtain the same coefficients combinations on the right-hand side as for the parallel case, except for the source term, with a matching for $N=1$. However, this linear combination does not allow to conclude either.

In the other linear combination considered, we sum only up to $N-1$, with $N=4$ as in known solutions. A motivation for this is the independence of the scalar fields that we have not discussed so far. As long as $N\leq 5$, $\rho$ and the $\sigma_{I}$ define at most six (diagonal) metric fluctuations, which seem a priori independent. However, due to the specific source configuration considered here, namely the homogeneous overlap \eqref{homoov}, there is a redundancy as we now show. Disregarding the group manifold algebra, all internal directions are a priori equivalent, so we can have the set $I=1$ along $123$. Then the homogeneous overlap with $N_o=1$ imposes the other three sets to have one direction along $123$ and two directions along $456$. Up to $\mathbb{Z}_3$ permutations on the first direction, we can fix $I=2$ to be along $145$, $I=3$ along $256$ and $I=4$ along $346$. This gives the following dependence on metric fluctuations
\bea
& e^1= \sqrt{\rho \sigma_1^A \sigma_2^A \sigma_3^B \sigma_4^B}\ e^{1\, 0} \ ,\ e^2= \sqrt{\rho \sigma_1^A \sigma_3^A \sigma_2^B \sigma_4^B}\ e^{2\, 0} \ ,\  e^3= \sqrt{\rho \sigma_1^A \sigma_4^A \sigma_2^B \sigma_3^B}\ e^{3\, 0} \ ,\\
& e^4= \sqrt{\rho \sigma_2^A \sigma_4^A \sigma_1^B \sigma_3^B}\ e^{4\, 0} \ ,\ e^5= \sqrt{\rho \sigma_2^A \sigma_3^A \sigma_1^B \sigma_4^B}\ e^{5\, 0} \ ,\  e^6= \sqrt{\rho \sigma_3^A \sigma_4^A \sigma_1^B \sigma_2^B}\ e^{6\, 0} \ .\nn
\eea
One verifies that setting $\sigma_{I=4}$ to $1$, it can be recovered by the rescaling
\beq
\rho \rightarrow \rho\ \sigma_4^{2(A+B)} \ ,\ \sigma_{I\neq 4} \rightarrow \sigma_{I\neq 4}\ \sigma_4^{-1} \ ,
\eeq
so this fluctuation is redundant. We cannot get rid of more scalars: for instance, setting $\sigma_{I=4}=1$, the dependence on $\sigma_1$ is different in $e^1$ and $e^4$ while that on the remaining fields is the same, so $\sigma_1$ could not be reobtained by rescaling the other fields. To conclude, the derivation of the potential regarding each $\sigma_I$ is correct, but we can pick a gauge where $\sigma_{I=4}=1$. The corresponding source is still there, so $\delta_{a_{||_J}}^{a_{||_I}}$ and $N_o$ still make sense, $T_{10}^4$ and its Bianchi identity as well. Considering the corresponding linear combination where we sum only up to $N-1=3$ is however not more successful: we do not reach conclusive results.\\

Let us now compare to two papers \cite{Danielsson:2012et, Junghans:2016uvg} where conjecture \ref{conj2} has been verified on explicit examples, namely on the SU(2)$\times$SU(2) group manifold, using similar tools. In \cite{Junghans:2016uvg}, all e.o.m. are considered, and the derivation depends on the precise solution ansatz. In \cite{Danielsson:2012et}, the tachyonic scalar potential is given after taking into account two further extremum conditions, corresponding to $C_3$ axions, so presumably related to the $F_4$-flux e.o.m. In \cite{Garg:2018zdg}, the same potential is used and a tachyon is obtained numerically when using information from a $B$-field axion, so probably related to the $H$-flux e.o.m. The conclusion appears to be similar to what is presented at the beginning of section \ref{sec:conj1} for conjecture \ref{conj1}: we need to include information coming from other equations, not considered so far. As explained already, this is however difficult because remaining equations do not combine easily with the above linear combinations, since their entries are of a different kind. We still derived new interesting constraints in section \ref{sec:existinter}, but those remain too weak or not appropriate. So for now, the strategy and tools developed here do not allow to conclude on conjecture \ref{conj2}, but if some extra information is provided, such as a flux ansatz or the algebra of the group manifold, or even some useful combination of remaining equations, then the machinery developed here could turn out to be conclusive.

\subsection{Further existence constraints}\label{sec:existinterfurther}

There is a special case of intersecting sources for which the above machinery provides a natural extension of existence constraints obtained for parallel sources in \cite{Andriot:2018ept}, and discussed in section \ref{sec:c}. As one intuitively understands, it is the case where one set of parallel sources is dominant with respect to other sets. The only difference in the general linear combination \eqref{galLC3} between the parallel and intersecting cases is the term in $\sum_{J\neq I} T_{10}^J$: when this term has a proper sign, the constraints derived in \cite{Andriot:2018ept} can then be generalized. As we will see, the sign is $\sum_{J\neq I} T_{10}^J = T_{10} - T_{10}^I \leq 0$, meaning that the sets $J\neq I$ contain overall few or no orientifold, i.e.~are dominated by brane contributions, while the set $I$ has mostly orientifold contributions as in a parallel case.

We focus on the general linear combination \eqref{galLC3}, staying more general than in section \ref{sec:conj2attempt} by allowing $p=4,5,6$ and $N_o$ unfixed. We take specific coefficients from \cite{Andriot:2018ept} with $A=p-9$, $B=p-3$, namely
\beq
b_{\tau}= \frac{3}{2}a\ ,\ b_{\rho}= \frac{A+B}{A-B} a\ ,\ b_{\sigma}= \frac{2}{B-A} a \ ,\ d = 4a \ ,\  c_{\tau}= c_{\rho} = c_{\sigma} =0 \ ,\label{coefl0}
\eeq
which give
\bea
& {\cal R}_4 + \frac{3}{2} \ \tau \del_{\tau} \tV|_0  + \frac{A+B}{A-B} \ \rho \del_{\rho} \tV|_0  + \frac{2}{B-A} \ \sigma_I \del_{\sigma_I} \tV|_0  \label{combil0}\\
& +2  \left|*_{\bot_I}H^{(0)_I} + \varepsilon_p g_s F_{k-2}^{(0)_I} \right|^2  + 2 \sum_{a_{||_I}} \left| *_{\bot_I}( \d e^{a_{||_I}})|_{\bot_I} - \varepsilon_p g_s\, \iota_{a_{||_I}} F_k^{(1)_I} \right|^2 \nn\\
= & -2 g_s^2 \left( |F_{k+2}|^2 + |F_{k+4}|^2 \right) - 2\, \delta^{cd}   f^{b_{\bot_I}}{}_{a_{||_I}  c_{\bot_I}} f^{a_{||_I}}{}_{ b_{\bot_I} d_{\bot_I}} + g_s \sum_{J\neq I} \frac{T_{10}^J}{p+1} (N_o + 7-p)  \ .\nn
\eea
The last term is negative in the case $\sum_{J\neq I} T_{10}^J \leq 0 $, which will allow us to conclude as in the parallel case. We then take the following coefficients
\beq
b_{\tau} = \frac{A-5B}{A-3B} \frac{a}{2} \ ,\ b_{\rho}= \frac{a}{3} \ ,\ b_{\sigma} = \frac{2}{3(A-3B)}a \ ,\  c_{\tau}= c_{\rho} = c_{\sigma} = d=0 \ ,\label{coefl1}
\eeq
which give
\bea
& {\cal R}_4 + \frac{1}{2} \frac{A-5B}{A-3B} \ \tau \del_{\tau} \tV|_0  + \frac{1}{3} \ \rho \del_{\rho} \tV|_0  + \frac{2}{3(A-3B)} \ \sigma_I \del_{\sigma_I} \tV|_0 \label{combil1}\\
= & -\frac{2}{p} \left( - \delta^{cd}   f^{b_{\bot_I}}{}_{a_{||_I}  c_{\bot_I}} f^{a_{||_I}}{}_{ b_{\bot_I} d_{\bot_I}} -  |f^{{}_{||_I}}{}_{{}_{\bot_I} {}_{\bot_I}}|^2  + |H^{(2)_I}|^2 \right) \nn\\
& - \frac{2}{p}\, g_s^2 \frac{1}{2} \sum_{q=6-p}^p (q - (6- p))\ |F_q|^2   + \frac{1}{p} g_s \sum_{J\neq I} \frac{T_{10}^J}{p+1}  (p-3 -N_o)  \nn \ ,
\eea
where again the last term is negative for $\sum_{J\neq I} T_{10}^J \leq 0 $. We now assume the latter holds. If $|f^{{}_{||_I}}{}_{{}_{\bot_I} {}_{\bot_I}}|^2 = 0$, any such structure constant vanishes, so $\delta^{cd}   f^{b_{\bot_I}}{}_{a_{||_I}  c_{\bot_I}} f^{a_{||_I}}{}_{ b_{\bot_I} d_{\bot_I}} = 0$. The right-hand side of \eqref{combil1} is then negative, while the left-hand side should be strictly positive, so we reach a no-go: for $\sum_{J\neq I} T_{10}^J \leq 0 $, one must have $|f^{{}_{||_I}}{}_{{}_{\bot_I} {}_{\bot_I}}|^2 \neq 0$. One can then safely define a parameter $\lambda_I$ (see section \ref{sec:existinter}) as
\beq
\delta^{cd}   f^{b_{\bot_I}}{}_{a_{||_I}  c_{\bot_I}} f^{a_{||_I}}{}_{ b_{\bot_I} d_{\bot_I}} = - \lambda_I |f^{{}_{||_I}}{}_{{}_{\bot_I} {}_{\bot_I}}|^2 \ . \label{deflambdaI}
\eeq
We deduce from \eqref{combil0} and \eqref{combil1} the following existence requirements
\beq
\mbox{de Sitter}:\quad {\rm if}\ \exists I\ \mbox{s.t.}\ \sum_{J\neq I} T_{10}^J \leq 0,\ {\rm then}\ |f^{{}_{||_I}}{}_{{}_{\bot_I} {}_{\bot_I}}|^2 \neq 0 \  \ {\rm and}\ \ 0 < \lambda_I < 1 \ .\label{constraint}
\eeq
This is valid on group manifolds with constant fluxes, $p=4,5,6$, and at least for homogeneous overlap \eqref{homoov}. As in the parallel case, this gives constraints on the type of group manifolds and underlying algebras that are allowed, excluding in particular nilmanifolds in their standard basis, where $\lambda_I = 0$; we refer to \cite{Andriot:2018ept} for more detail on these constraints.

As discussed in section \ref{sec:existinter}, one may then generalize to the intersecting case $\sum_{J\neq I} T_{10}^J \leq 0$ the new constraints obtained in section \ref{sec:othereqpar} in the parallel case from the Einstein equation. More generally, to go further with the contribution $\sum_{J\neq I} T_{10}^J $ (especially for known solutions that allow a priori the same orientifold contribution along each set), one may consider the following linear combination
\bea
& a {\cal R}_4 + b_{\tau} \ \tau \del_{\tau} \tV|_0  + b_{\rho} \ \rho \del_{\rho} \tV|_0  + b_{\sigma} \ \sigma_I \del_{\sigma_I} \tV|_0 \label{galLC9}\\
& + c_{\tau} \ \tau^2 \del^2_{\tau} \tV|_0 + c_{\rho} \ \rho^2 \del^2_{\rho} \tV|_0 + c_{\sigma} \ \left(  \sigma_I^2 \del^2_{\sigma_I} \tV|_0 + \sigma_I \del_{\sigma_I} \tV|_0 \right)\nn\\
& +\frac{d_I}{2} \left|*_{\bot_I}H^{(0)_I} + \varepsilon_p g_s F_{k-2}^{(0)_I} \right|^2  + \frac{d_I}{2} \sum_{a_{||_I}} \left| *_{\bot_I}( \d e^{a_{||_I}})|_{\bot_I} - \varepsilon_p g_s\, \iota_{a_{||_I}} F_k^{(1)_I} \right|^2 \nn\\
& +\frac{d_J}{2} \sum_{J\neq I} \left|*_{\bot_J}H^{(0)_J} + \varepsilon_p g_s F_{k-2}^{(0)_J} \right|^2  + \frac{d_J}{2} \sum_{J\neq I} \sum_{a_{||_J}} \left| *_{\bot_J}( \d e^{a_{||_J}})|_{\bot_J} - \varepsilon_p g_s\, \iota_{a_{||_J}} F_k^{(1)_J} \right|^2 \ ,\nn
\eea
which includes contributions of the $F_k$ BI projected on the sets $J$. Analysing it would require to compare structure constants and fluxes projected along the different sets, an involved task that we leave to future work.

\section{Conjecture 3: no supergravity solution with string origin}\label{sec:conj3}

As explained in the introduction, 10d supergravity is a low energy effective theory for string theory provided some conditions are verified. The energy approximation requires any length to be much bigger than the string length, $L \gg l_s$, to neglect string states as well as $\alpha'$ or higher derivative corrections. The string coupling constant should be small to neglect string loop corrections, here given by $g_s \ll 1$. These requirements ensure a low energy, classical regime, of string theory. Another requirement is that classical solutions of extended supergravity object correspond to stringy objects, here $O_p$ and $D_p$. For this to hold, their number has to be quantized, and for orientifolds, it is bounded from above because it corresponds to fixed points of the geometry (e.g.~two per circle). Finally, fluxes should also be quantized, at least on cycles of the geometry; one reason is cohomology. This quantization however involves $L$ and $l_s$ so a quantized flux gets ``diluted'' when the former is bigger than the latter, i.e.~in a large volume limit. Having a solution of supergravity is thus a first result, making it a classical string background asks for these further requirements to be satisfied. This idea was recently put forward in \cite{Roupec:2018mbn} when analysing known classical de Sitter solutions of type IIA supergravity: those appeared inconsistent with a stringy origin, because of these extra requirements. This question was then studied formally and exemplified for $p=6$ in \cite{Junghans:2018gdb, Banlaki:2018ayh}: the conclusions turned out consistent with conjecture \ref{conj3}, without being fully general. We investigate this question here for $p=4,5,6$ in a 10d approach and argue that one cannot conclude on a proof of conjecture \ref{conj3} with the tools developed so far. The case of a compact manifold with an internal hierarchy appears indeed consistent with the various requirements, and in our view, cannot be excluded it for now.\\

We first give the tools necessary to this study. For a flux $F_q$ on a cycle $\Sigma_q$, the quantization goes as
\beq
\frac{1}{(2\pi {\alpha'}^{\tfrac{1}{2}})^{q-1}} \int_{\Sigma_q} F_q = \frac{1}{(2\pi l_s)^{q-1}} \int_{\Sigma_q} {\rm vol}_q\, F_{q\, a_1 \dots a_q} = N_q \in \mathbb{Z} \ ,
\eeq
where the volume is extracted by writing the flux component in the flat (orthonormal) basis of the metric. With our ansatz, these components are constant; otherwise one can consider the integral of the component divided by the volume. The component quantization is then given by $F_{q\, a_1 \dots a_q} = N_q (2\pi l_s)^{q-1} / (\int_{\Sigma_q} {\rm vol}_q)$. Writing the volume as an average $(2\pi L)^q$ and absorbing the $2\pi$ in the lengths, we deduce
\beq
|F_q|^2 = \frac{N_q^2}{l_s^2} \left(\frac{l_s}{L}\right)^{2q} \ .
\eeq
This may look too schematic, due to possible caveats. First, on group manifolds, it is not clear one always has the appropriate cycles. Also, the last square involves a sum on several components, so $N_q^2$ may rather denote a sum of squared integers. Finally, an average length $L$, even though sufficient when comparing internal lengths to $l_s$, might be too restrictive for a refined analysis that would include internal hierarchies. For now, we stick to this schematic or restrictive approach and leave these caveats aside, it will be enough for our study. Note that we reproduce the 4d potential approach: the volume field $\rho$ is replaced by $L$ and $\tau^{-1}$ by $g_s$.

We turn to source contributions: $T_{10}^I$ is defined in \cite{Andriot:2017jhf} in terms of charges and $\delta$-functions, and we take an integrated version as prescribed by the smeared ansatz. We obtain
\beq
\frac{T_{10}^I}{p+1}= (2^{p-5} N^I_{O_p} - N^I_{D_p})\, \frac{(2\pi l_s)^{7-p}}{\sqrt{|g_{\bot_I}|}} = \frac{N_s^I}{l_s^2} \left(\frac{l_s}{L}\right)^{9-p} \ ,
\eeq
where $N_s^I = 2^{p-5} N^I_{O_p} - N^I_{D_p}$ is the ``number of sources'' given in terms of the number of orientifolds $N^I_{O_p}$ and branes $N^I_{D_p}$ in the set $I$, and $\sqrt{|g_{\bot_I}|}$ denotes the transverse volume to this set, expressed again with an average internal length where $2\pi$ was absorbed. $N_s^I$ can be considered as an integer, bounded from above because of $N^I_{O_p}$. Summing $T_{10}= \sum_I T_{10}^I$, we further write
\beq
\frac{T_{10}}{p+1}= \frac{N_s}{l_s^2} \left(\frac{l_s}{L}\right)^{9-p} \ ,
\eeq
where the resulting number $N_s$ is also bounded from above. It is a finite, not very big, number, that has to be strictly positive for a de Sitter solution.

We finally focus on the internal curvature ${\cal R}_6$: it is given on group manifolds by a linear combination of squares of structure constants, as in \eqref{Ricci}, \eqref{R6gp}. The structure constants are discretized due to the compactness of the manifold, which can be seen through the lattice action. For instance, the nilmanifold based on the Heisenberg Lie algebra has the structure constant
\beq
f^3{}_{12}= \frac{r^3}{r^1 r^2} N_{f}^i \ ,\label{fnil}
\eeq
where $N_{f}^i$ is an integer and $r^a$ are the radii. More generally, in our conventions, the structure constants include lengths in the same manner, and they are proportional to a number $N_f$ that is discretized. Ignoring internal length hierarchies and using an average length, we then write schematically
\beq
{\cal R}_6 = - \frac{N_f^2}{L^2} \ ,\label{R6L}
\eeq
since we know that ${\cal R}_6$ has to be strictly negative for a de Sitter solution. Considering however $N_f^2$ in \eqref{R6L} to be an integer is a priori too strong. It is defined as the resulting non-zero number obtained when extracting the average length, and as we will see, it could actually be very small, once one takes into account internal hierarchies or because of the special expression of ${\cal R}_6$.\\

We now start the analysis and study the limits $l_s/L \rightarrow 0$, $g_s \rightarrow 0$ or corresponding regimes, with $N_s$ bounded. We combine appropriately the equations \eqref{eqR4}, \eqref{eqdeltauV}, \eqref{eqdelrhoV}, to be satisfied by de Sitter solutions. We read from \eqref{eqdeltauV}, using that $F_{6-p} \neq 0$,
\beq
3 g_s \frac{T_{10}}{p+1} > - 2 {\cal R}_6  \quad \Rightarrow \quad  N_s > \frac{2}{3} \frac{N_f^2}{g_s} \left(\frac{L}{l_s}\right)^{7-p} \ . \label{T10R6}
\eeq
It is tempting to conclude from \eqref{T10R6} a proof of conjecture \ref{conj3}: indeed, taking $N_f^2 \geq 1$, the requirements  $L \gg l_s$ and $g_s \ll 1$ lead to a large right-hand side since $p=4,5,6$, forcing $N_s$ to be larger than its bound; this point was made already for $p=6$ in \cite{Junghans:2018gdb, Banlaki:2018ayh}. For instance, a hierarchy of order 10 gives already $N_s$ of at least 100, which is above the standard value of $2^{9-p}$. The caveat in this reasoning appears when allowing for internal hierarchies, i.e.~going beyond the use of an average length and considering the possibility of $N_f^2$ being a small number. This leads us to conclude
\begin{empheq}[innerbox=\fbox]{align}
& \mbox{Conjecture \ref{conj3} holds unless the internal space} \label{hierarchystatement}\\
& \mbox{admits hierarchies such that}\ N_f^2= |{\cal R}_6|\times L^2 \ll 1 \ . \nn
\end{empheq}
There are two ways to get a small $N_f^2$. First, $L$ is here the average internal length, which can be viewed as fixed by the volume. When rather considering the radii separately and allowing for internal hierarchies, the curvature ${\cal R}_6$ can generate a different length scale. For instance, this not the case in isotropic spheres, but it can easily be achieved in nilmanifolds: as can be seen in \eqref{fnil}, taking the fiber radius small compared to the base ones, $r^3 \ll r^1, r^2$, generates a hierarchy, eventually leading to a small $N_f^2$.\footnote{For $\mmm= {\rm Nil}_3 \times T^3$, one has ${\cal R}_6= - \tfrac{1}{2} (f^3{}_{12})^2$. When all radii except $r^3$ are equal to $R$, one computes $N_f^2 = (r^3/R)^{\frac{7}{3}}\, {N_{f}^i}^2 / 2 $.} Secondly, ${\cal R}_6$ is a linear combination of squares of structure constants, where positive and negative contributions could be tuned and almost cancel each other, giving eventually a small but non-vanishing value. Using the general expression \eqref{R6gp} of ${\cal R}_6$ on group manifolds with the parameter $\lambda$ \eqref{lambda} (see \eqref{R6group}), we could not bound $-{\cal R}_6$ from below by a single positive term, i.e.~it could be tuned as close to zero as desired. Even though a detailed analysis of each component of Einstein equation may forbid this, we can for now not exclude an arbitrarily small $N_f^2$.

Having such an internal hierarchy was considered to some extent in \cite{Junghans:2018gdb}, when studying the possibility of a classical de Sitter solution in a controlled regime in the limit of a large field $\alpha$. Interestingly, this analysis was very general as it allowed $\alpha$ to be any field combination. One could then imagine $\alpha$ being the right combination to parametrize a small $N_f^2$, i.e.~our internal hierarchy. The analysis concluded on the general difficulty to have a classical de Sitter solution in a large field limit, whatever $\alpha$ is. This is in line with the discussion of \cite{Ooguri:2018wrx} and the Dine--Seiberg argument \cite{Dine:1985he}. This would exclude the possibility of our internal hierarchy. Several possible loopholes were however mentioned in \cite{Junghans:2018gdb}. The most relevant one to us is the fact that, even on a group manifold, ${\cal R}_6$ could be a complicated function of the various fields, without a fixed sign, and its dependence on the appropriate $\alpha$ is hard to predict in our view. More generally, studying a limit or an asymptotic regime {\it {\`a} la} Dine--Seiberg, where e.g.~one term in a potential dominates the others, could be misleading here. Indeed, the de Sitter solution could be located in a grey zone, between ``the interior of field space'' and the asymptotics, where some fields are large enough to give a controlled regime, but not so large as to approximate each appearing function by a single term. This relates to the idea of ${\cal R}_6$ being a complicated function whose behavior is difficult to capture. These options leave some freedom to have the appropriate internal hierarchy \eqref{hierarchystatement}, so we cannot exclude this possibility for now.\\

Having compared the source term to the curvature term, we turn to the fluxes. With them we show, in some conditions at least, that one cannot take independently the limits $l_s/L \rightarrow 0$, $g_s \rightarrow 0$, rather the two are actually related. From \eqref{R4T10F}, one deduces
\beq
g_s \frac{T_{10}}{p+1} > g_s^2\, |F_{6-p}|^2 \quad \Rightarrow \quad N_s > N_{6-p}^2\ g_s \left(\frac{L}{l_s}\right)^{p-3} \ .
\eeq
In addition, combining \eqref{eqdeltauV} $- 2\times$ \eqref{eqdelrhoV} together with \eqref{R4T10F}, one deduces for de Sitter
\beq
\frac{8-p}{2}\, g_s \frac{T_{10}}{p+1} > |H|^2 \quad \Rightarrow \quad \frac{8-p}{2}\, N_s > N_{H}^2\ \frac{1}{g_s} \left(\frac{l_s}{L}\right)^{p-3} \ .\label{boundH}
\eeq
All flux terms are bounded by the source term (all RR fluxes are bounded with \eqref{R4T10F}): this forbids having too big numbers $N_q$ of fluxes. In addition, the quantity $g_s (L/l_s)^{p-3}$, undetermined in the supergravity regime, cannot be either too big or too small. In the case where $N_{6-p}$ and $N_{H}$ are integers bigger or equal to 1, the quantity gets bounded as follows
\beq
\frac{2}{8-p}\ \frac{1}{N_s} < g_s \left(\frac{L}{l_s}\right)^{p-3} < N_s \ , \label{ineqNs}
\eeq
which relates the two limits: indeed, for a reasonable $N_s \sim O(1)$, we deduce $g_s \sim (l_s/L)^{p-3}$ and $N_f^2 \lesssim (l_s/L)^4$, the latter indicating a strong internal hierarchy.\footnote{In \cite{Banlaki:2018ayh} was proposed for $p=6$ the scaling $L/l_s \sim g_s^{-1} \sim \sqrt{N_s}$ for large $N_s$. This is different than considering as here a finite $N_s$ of order $1$. Still, if $N_s$ had to be large, the inequality \eqref{ineqNs} is allowing the relation $g_s (L/l_s)^{p-3} \sim N_s$, consistent with the scaling of \cite{Banlaki:2018ayh}.} With such values, every requirement and equations considered so far would be consistent. Note that the condition $N_{6-p}, N_{H} \geq 1$ is not generic for two reasons. First, among the two fluxes, only $F_{6-p}$ is a priori required to be non-zero. Secondly, in the case of internal hierarchies, the difference with the average length could give different values to the numbers $N_{6-p}, N_{H}$. In that case, one could still derive analogous inequalities, even though possibly less constraining. In addition, in the parallel case with $p=6$ and $F_4=F_6=0$, we reach such inequalities without any assumption. Indeed we showed in \eqref{Hneq0par} that $H^{(0)} \neq 0$, and one can rewrite \eqref{boundH} with $|H|^2 \geq |H^{(0)}|^2$; furthermore the lengths involved in the inequalities are then precisely the same, namely the transverse volume, and one relates eventually the small $g_s$ to the large transverse volume limit. There is thus a chance to get constraining inequalities \eqref{ineqNs} quite generally, and relate this way the two limits.

Finally, using \eqref{boundH} with \eqref{eqR4} and \eqref{R4T10F}, one reaches
\beq
{\cal R}_4 < - {\cal R}_6 + \frac{6-p}{4} g_s \frac{T_{10}}{p+1} \quad \Rightarrow \quad \left(\frac{L}{L_{4d}}\right)^2 < N_f^2 + \frac{6-p}{4} N_s g_s  \left(\frac{l_s}{L}\right)^{7-p} \ ,
\eeq
where ${\cal R}_4 = 1/L_{4d}^2$. The required small $N_f^2$ and the appropriate supergravity regime imply a scale separation with 4d, $L/L_{4d} \ll 1$: this would certainly be an interesting situation.\footnote{While the equations derived here agree with those of \cite{Gautason:2015tig} (for instance for $p=6$, we get $|{\cal R}_6 / {\cal R}_4| < O(1) $ as there), the interpretation of the result differs. We consider here that the separation of scale should be defined by a large hierarchy between the 4d scale and the internal typical scale $L$, defined e.g.~by the volume; we have in mind that the internal Kaluza--Klein energy scale, defined by the (square root of the) first eigenvalue of the Laplacian on scalar fields, should be close to $1/L$. On the contrary, the internal scale compared to the 4d scale in \cite{Gautason:2015tig} is defined by the average of $|{\cal R}_6|$; the authors also specify that if this scale is different than the Kaluza--Klein scale, their analysis does not apply. We believe that for manifolds admitting the desired internal hierarchy, i.e.~having a small $N_f^2$, both the first scalar eigenvalue and $1/L^2$ are very different than the curvature scale fixed by $|{\cal R}_6|$. An example is the explicit analysis of the Laplacian spectrum on the nilmanifold \eqref{fnil} in \cite{Andriot:2016rdd, Andriot:2018tmb}: in a regime of ``small fiber'', one generates the desired hierarchy, where indeed $|{\cal R}_6|$ is small compared to the first scalar eigenvalue and to $1/L^2$.} The question remains whether the necessary internal hierarchy can be reached with a de Sitter solution.\\

We did not find further useful information studying the flux equations, in particular the $F_k$ BI, the $H$-flux e.o.m. and the new results and relations derived in sections \ref{sec:othereqpar} or \ref{sec:existinter}. For instance, for $p=6$, all lengths drop out of the $F_2$ BI projected on the transverse space, which can then be written schematically
\beq
N_{f^{{}_{||}}{}_{{}_{\bot} {}_{\bot}}} N_2 + N_{H^{(0)}} N_0 \sim N_s \ .
\eeq
Thanks to the results \eqref{signHF},  \eqref{dF2pos}, and \eqref{dF2posinter}, one may argue that the two terms on the left-hand side have the same sign, i.e.~do not compete against each other. This implies that they are bounded. But this does not contradict any of the previous requirements. In particular, the number $N_{f^{{}_{||}}{}_{{}_{\bot} {}_{\bot}}}$ related to the structure constants $f^{{}_{||}}{}_{{}_{\bot} {}_{\bot}}$ is not necessarily an integer. It can either be small and generate an internal hierarchy, or it can be e.g.~of order $N_s$ and compensated in ${\cal R}_6$ by contributions of the other two structure constants listed in \eqref{fabcOp} to generate a small $N_f^2$, as explained below \eqref{hierarchystatement}. These open possibilities are to be contrasted with the discussion of section 2.3 of \cite{Banlaki:2018ayh}. The other flux equations do not allow us to be more conclusive regarding conjecture \ref{conj3}, because the same possibilities appear with structure constants, consistently with the required hierarchies. A detailed analysis of Einstein equation seems necessary to determine whether such internal hierarchies can occur in de Sitter solutions or not.

\section{Results summary}\label{sec:ccl}

In this paper we have studied three conjectures on classical de Sitter solutions, defined in the introduction. We have restricted our analysis to a common solution ansatz, for which proving any of these conjectures is a well-defined mathematical problem, as detailed in section \ref{sec:defpb}. We obtained the following results.
\begin{itemize}
  \item {\bf Conjecture \ref{conj1}}: we first considered in section \ref{sec:c} recent existence constraints on classical de Sitter solutions with parallel sources \cite{Andriot:2018ept}, and computed for those the number $c$ entering the swampland conjecture \cite{Obied:2018sgi}. This required to determine the kinetic term for the scalar field $\sigma$. We discussed further existence conditions, and obtained in section \ref{sec:othereqpar} new ones, on structure constants in \eqref{reqRpar}, \eqref{requirefabc}, \eqref{reqRbot}, \eqref{reqfbot}, and on fluxes in \eqref{Hneq0par}, \eqref{signHF}, \eqref{dF2pos}. We finally studied in section \ref{sec:partcase} a particular case, where all equations could be solved consistently with a de Sitter solution, except a few internal components of Einstein equation that were difficult to tackle generically. This proves that the detailed information of Einstein equation (and not just some of its traces) is needed to prove conjecture \ref{conj1}, which makes it more challenging.
  \item {\bf Conjecture \ref{conj2}}: we reviewed in section \ref{sec:existinter} the existence conditions for classical de Sitter solutions with intersecting sources, and found new ones on fluxes in \eqref{Hneq0inter}, \eqref{signHFinter}, \eqref{dF2posinter}. We then determined in section \ref{sec:pot} the scalar potential \eqref{pot} of the scalar fields $\rho, \tau, \sigma_{I=1\dots N}$ and its derivatives, generalizing to intersecting sources the work done in \cite{Andriot:2018ept} on parallel sources. The strategy presented was then applied in section \ref{sec:conj2attempt} to the case of known solutions, but we failed to prove the presence of a systematic tachyon. Doing so seemed to require combining our new formalism with more information coming from other equations. Nevertheless, the tools developed allowed us to obtain further existence constraints in \eqref{constraint}, and another suggestion \eqref{galLC9} for the tachyon.
  \item {\bf Conjecture \ref{conj3}}: we first showed, for $p = 4, 5, 6$, that this conjecture would hold unless the compact manifold admits a hierarchy such that $|{\cal R}_6| \times L^2 \ll 1$, where $L$ is an average internal length \eqref{hierarchystatement}. We were not able to exclude the possibility of such an internal hierarchy in a classical de Sitter solution, despite interesting arguments of \cite{Junghans:2018gdb}, and detailed information from the Einstein equation seemed needed to that end. Interestingly, we showed that having this hierarchy would lead to a scale separation with 4d. Finally, we proved that the desired limits $l_s/L \rightarrow 0$ and $g_s \rightarrow 0$ could not be independent, at least for $H \neq 0$ and further conditions, but are related or bounded as in \eqref{ineqNs}.
\end{itemize}
We hope that this work will trigger more thorough analysis in the remaining corners of parameter space where these conjectures are not proven. In particular, the case of section \ref{sec:partcase} deserves more attention.

\vspace{0.4in}

\subsection*{Acknowledgements}

We are indebted to D.~Junghans, C.~Roupec, T.~Van Riet and T.~Wrase for helpful discussions.

\newpage

\begin{appendix}

\section{Type II supergravities equations}\label{ap:eq}

We read from \cite{Andriot:2016xvq} the 10d Einstein equation and its trace, from which we get the trace-inversed Einstein equation
\bea
{\cal R}_{MN} & = \frac{1}{4} H_{MPQ}H_N^{\ \ PQ}+\frac{e^{2\phi}}{2}\left(F_{2\ MP}F_{2\ N}^{\ \ \ \ P} +\frac{1}{3!} F^{10}_{4\ MPQR}F_{4\ N}^{10 \ \ PQR} \right) \label{EinstIIAr}\\
& + \frac{e^{\phi}}{2}T_{MN} -2\nabla_M \del_N{\phi} \nn\\
& + \frac{g_{MN}}{16} \left( - e^{\phi} T_{10} - 2|H|^2 + e^{2\phi}(|F_0|^2 - |F_2|^2 -3 |F_4|^2 + 3 |F_6|^2 ) - 4 \Delta \phi +8 |\del \phi|^2 \right) \nn \ , \\
{\cal R}_{MN} & = \frac{1}{4} H_{MPQ}H_N^{\ \ PQ}+\frac{e^{2\phi}}{2}\left(F_{1\ M}F_{1\ N} +\frac{1}{2!} F_{3\ MPQ}F_{3\ N}^{\ \ \ \ PQ} +\frac{1}{2\cdot4!} F_{5\ MPQRS}^{10}F_{5\ N}^{10 \ \ PQRS} \right) \nn\\
& + \frac{e^{\phi}}{2}T_{MN} -2\nabla_M \del_N{\phi} \label{EinstIIBr}\\
& + \frac{g_{MN}}{16} \left( - e^{\phi} T_{10} - 2|H|^2 - 2 e^{2\phi} |F_3|^2 - 4 \Delta \phi +8 |\del \phi|^2 \right) \nn \ ,
\eea
where anticipating on the compactification ansatz we used $|F_4^{10}|^2 = |F_4|^2 - |F_6|^2 $. We denote $|A_q|^2 = A_{q\, M_1 \dots M_q}\, g^{M_1N_1} \! \dots g^{M_q N_q} A_{q\, N_1 \dots N_q} / q!$ for a $q$-form $A_q$. We express the source contribution $T_{MN}$ using flat indices: $T_{AB}= e^M{}_A e^N{}_B T_{MN}$, with the decomposition into the 4d flat directions $\alpha$, the 6d $a_{||}$ and $a_{\bot}$ for each source. Following \cite{Andriot:2017jhf}, and using the sources ansatz described in section \ref{sec:ansatz}, we get for each source $T_{a_{\bot}b_{\bot}} = e^M{}_{a_{\bot}} e^N{}_{b_{\bot}} T_{MN}=0$ and
\beq
T_{AB}= \delta_A^{\alpha} \delta_B^{\beta}\, T_{\alpha \beta} + \sum_I \delta_A^{a_{||_I}} \delta_B^{b_{||_I}} \, T^I_{a_{||_I}b_{||_I}} \ , \label{T1}
\eeq
with the sum over the sets $I=1 \dots N$ of intersecting sources; in the case of parallel sources, $N=1$ so the sum has only one term and one drops the index $I$. The trace $T_{10}=g^{MN} T_{MN}$ gets decomposed similarly $T_{10} = \sum_{I} T_{10}^I$, and eventually, one obtains the expressions
\beq
T_{\alpha \beta}= \eta_{\alpha \beta} \frac{T_{10}}{p+1}  \ , \quad T^I_{a_{||_I}b_{||_I}} = \delta_{a_{||_I}b_{||_I}} \frac{T_{10}^I}{p+1}  \ .\label{T2}
\eeq
Source contributions are thus only expressed in terms of the numbers $T_{10}^I$ and their sum $T_{10}$, together with overlap numbers $\delta_A^{a_{||_I}} $.

With the flux compactification ansatz \cite{Andriot:2016xvq}, we obtain the following 4d components
\beq
\frac{1}{3!} F^{10}_{4\ \mu PQR}F_{4\ \nu}^{10 \ \ PQR} = - g_{\mu \nu} |F_6|^2 \ ,\quad \frac{1}{4!} F_{5\ \mu PQRS}^{10}F_{5\ \nu}^{10 \ \ PQRS} = - g_{\mu \nu} |F_5|^2 \ .
\eeq
We deduce the 4d Einstein equation
\bea
\hspace{-0.4in} {\cal R}_{\mu \nu} & =  \frac{g_{\mu \nu}}{16} \left(e^{\phi} \frac{T_{10}}{p+1} (7-p) - 2|H|^2 + e^{2\phi}(|F_0|^2 - |F_2|^2 -3 |F_4|^2 -5 |F_6|^2 ) - 4 \Delta \phi +8 |\del \phi|^2 - 16 g^{pn} e^{-2A} \del_n e^{2A} \del_p \phi \right)  \ , \nn\\
\hspace{-0.4in} {\cal R}_{\mu \nu} & =  \frac{g_{\mu \nu}}{16} \left(e^{\phi} \frac{T_{10}}{p+1} (7-p) - 2|H|^2 - e^{2\phi}( 2 |F_3|^2 + 4 |F_5|^2 ) - 4 \Delta \phi +8 |\del \phi|^2 - 16 g^{pn} e^{-2A} \del_n e^{2A} \del_p \phi \right) \ ,\nn
\eea
where the last dilaton term comes from $\nabla_M \del_N{\phi}$ (see (C.3) and (C.4) of \cite{Andriot:2016xvq}). Since we consider a maximally symmetric 4d space-time, this equation is equivalent to its trace. We denote ${\cal R}_{10}= g^{MN} {\cal R}_{MN}$, ${\cal R}_4= g^{MN} {\cal R}_{MN=\mu\nu}$, ${\cal R}_6= g^{MN} {\cal R}_{MN=mn}={\cal R}_{10} - {\cal R}_{4}$. For completeness, we also give here the dilaton e.o.m.
\beq
2 {\cal R}_{10} + e^{\phi} \frac{T_{10}}{p+1} -|H|^2 + 8(\Delta \phi - |\del \phi|^2 ) = 0 \ . \label{dileom}
\eeq
This equation, together with the 4d and 10d traces of Einstein equation, are equivalent (for $A=0$ and constant $\phi$ at least) to the three 4d equations \eqref{eqR4}, \eqref{eqdeltauV}, \eqref{eqdelrhoV} \cite{Andriot:2018ept}.

\section{Kinetic terms}\label{ap:kin}

Kinetic terms of the scalars $\rho$ and $\tau$ are obtained in \cite{Hertzberg:2007wc}, with the same conventions as ours, and this derivation is justified by 4d supergravity arguments based on the K\"ahler potential. A more straightforward derivation could be made by a direct supergravity reduction, even though both are related. We propose here such a derivation following \cite{Roest:2004pk}, reproducing partially the results of \cite{Hertzberg:2007wc} and determining the kinetic term of the new scalar $\sigma$. We refer to \cite{Danielsson:2012et, Andriot:2018ept} for conventions regarding these three scalar fields.

The following standard result of dimensional reduction on group manifolds is given in \cite{Roest:2004pk} (section 3.4.2): if one considers the metric compactification ansatz
\beq
\d \hat{s}^2 = e^{2 \alpha \varphi} \d s^2 + e^{2 \beta \varphi} M_{ab} e^a e^b \ ,
\eeq
with $\det M = 1$, and $\alpha, \beta$ constants fixed according to dimensions, one gets the reduction
\beq
\sqrt{|\hat{g}|} \hat{{\cal R}} = \sqrt{|g|}  \left( {\cal R} + \frac{1}{4} {\rm Tr}(\del  M \del M^{-1}) - \frac{1}{2} (\del \varphi)^2 - V  \right) \ .\label{Lagterm}
\eeq
We use this result for our reduction, at least in the case where $\delta \phi = 0$, giving $\tau= \rho^{\frac{3}{2}}$. Indeed, spelling out our scalar fields \cite{Andriot:2018ept} gives the following metric
\beq
\d s_{10}^2 = \rho^{-3} \d s_{4E}^2 + \rho\, \d \tilde{s}_6^2 \ ,
\eeq
where one identifies $\d \tilde{s}_6^2 = M_{ab} e^a e^b $ and $\rho = e^{2\beta \varphi} = e^{-\frac{\varphi}{2 \sqrt{3}}}$ thanks to the values $\alpha^2 = \frac{3}{16}, \beta=-\frac{\alpha}{3}$, that provide a perfect match. The resulting Lagrangian term \eqref{Lagterm} becomes
\beq
\sqrt{|g_{4E}|}  \left( {\cal R}_{4E} + \frac{1}{4} {\rm Tr}(\del  M \del M^{-1}) - 4 (\del \hat{\rho})^2 - V  \right) \ ,
\eeq
using \eqref{canfields} for $\hat{\rho}$. This matches (up to an overall $\tfrac{1}{2}$) the following Lagrangian derived in \cite{Hertzberg:2007wc}, setting here $M_p=1$,
\beq
\frac{1}{2}  \left( {\cal R}_{4E} - (\del \hat{\rho})^2 - (\del \hat{\tau})^2 + \dots \right) \ ,
\eeq
using that here $\hat{\tau} = \sqrt{3} \hat{\rho}$, and where the dots include the other scalar fields.

This allows us to derive the kinetic term of $\sigma$. Importantly, this scalar field is defined such that it does not contribute to the 6d metric determinant, as required by $\det M = 1$: indeed we recall from \eqref{fluct} and \eqref{metricdet}
\beq
\d s_6^2 = \rho \left( \sigma^A (\d s_{||}^2)^0 + \sigma^B (\d s_{\bot}^2)^0 \right) \ , \quad A= p-9 \ ,\ B=p-3 \ .
\eeq
It is also the reason why $\sigma$ does not enter the definition of $\tau$. This allows to have independent (diagonal) kinetic terms. The matrix $M$ is given by diagonal blocks of $\sigma^A$ and $\sigma^B$, of size $p-3$ and $9-p$. One computes
\beq
{\rm Tr}(\del  M \del M^{-1}) = - B(9-p)(B-A)\ \sigma^{-2} (\del \sigma)^2 \ ,
\eeq
and we rewrite for convenience $B(9-p)(B-A) = - 6 AB$. We eventually deduce the following kinetic terms for our three scalars
\beq
\sqrt{|g_{4E}|}  \left( {\cal R}_{4E} - (\del \hat{\rho})^2 - (\del \hat{\tau})^2 - (\del \hat{\sigma})^2 \right) \ ,
\eeq
where we define
\beq
\hat{\sigma} = \sqrt{\frac{-3AB}{2}} \ln \sigma \ .
\eeq
A check of this canonically normalized field can be made with the explicit example discussed in \cite{Danielsson:2012et} (a choice is made there to keep only one $\sigma$), where one can read the scalar field metric from the K\"ahler potential: the same numerical value is then obtained.

\section{Scalar potential for intersecting sources}\label{ap:pot}

We derive the scalar potential discussed in section \ref{sec:pot} with intersecting sources, generalizing the case of parallel sources \cite{Andriot:2018ept}. To that end, we need to fluctuate the flux terms, the 6d curvature and the source terms. To get the contribution of (internal) fluxes, we decompose them along parallel and transverse directions to the sources: every flux component can be decomposed along $||_I$ and $\bot_I$ for any given $I$ as
\beq
F_q = \frac{1}{q!} F^{(0)_I}_{a_{1\bot_I} \dots a_{q\bot_I}} e^{a_{1\bot_I}} \w \dots \w e^{a_{q\bot_I}} + \frac{1}{(q-1)!} F^{(1)_I}_{a_{1||_I} a_{2\bot_I} \dots a_{q\bot_I}} e^{a_{1||_I}} \w e^{a_{2\bot_I}} \w \dots \w e^{a_{q\bot_I}} + \dots \nn
\eeq
and $|F_q|^2 = \sum_n |F_q^{(n)_I}|^2$ for any $I$. The dependence on $\sigma_I$ (and $\rho, \tau$) is then the same as in the parallel case. It is more complicated to express the dependence on $\sigma_{J \neq I}$ when having first developed along $||_I$ and $\bot_I$. To achieve this, one should know the details of the overlap of the different sets. We will not need to give any such explicit dependence on two different $\sigma_{I,J}$, so we only exhibit the dependence on one $\sigma_I$ and include that of the other $\sigma_J$ in parentheses. The same holds for the vielbeins and the 6d curvature: one can decompose the flat directions in $a_{||_I}, a_{\bot_I}$ for any $I$, and similarly, we just exhibit the dependence on one $\sigma_I$; we also denote all $\sigma_I$ as $\sigma_K$. We get
\bea
{\cal R}_6 (\sigma_K) = &\ \sigma_I^{-B} \left( {\cal R}_{\bot_I} +  \delta^{ab} \del_{a_{\bot_I}}  f^{c_{\bot_I}}{}_{c_{\bot_I}b_{\bot_I}} + {\cal R}_{\bot_I}^{||_I} + |f^{{}_{||_I}}{}_{{}_{\bot_I} {}_{\bot_I}}|^2 \right) (\sigma_J) \label{R6sigma}\\
& + \sigma_I^{-A} \left( {\cal R}_{||_I} + \delta^{ab} \del_{a_{||_I}}  f^{c_{||_I}}{}_{c_{||_I}b_{||_I}}  +  {\cal R}_{||_I}^{\bot_I} +|f^{{}_{\bot_I}}{}_{{}_{||_I} {}_{||_I}}|^2 \right)(\sigma_J) \nn\\
& - \frac{1}{2} \sigma_I^{-2A+B} |f^{{}_{\bot_I}}{}_{{}_{||_I} {}_{||_I}}|^2(\sigma_J) - \frac{1}{2} \sigma_I^{-2B+A} |f^{{}_{||_I}}{}_{{}_{\bot_I} {}_{\bot_I}}|^2(\sigma_J) \nn\\
= &\ \sigma_I^{-B}\ {\cal R}_6(\sigma_J) + (\sigma_I^{-A} - \sigma_I^{-B}) \left( {\cal R}_{||_I} + \delta^{ab} \del_{a_{||_I}}  f^{c_{||_I}}{}_{c_{||_I}b_{||_I}}  +  {\cal R}_{||_I}^{\bot_I} \right)(\sigma_J) \label{R6sigma3}\\
& - \frac{1}{2} (\sigma_I^{-2A+B} - 2\sigma_I^{-A} + \sigma_I^{-B}) |f^{{}_{\bot_I}}{}_{{}_{||_I} {}_{||_I}}|^2(\sigma_J) - \frac{1}{2} (\sigma_I^{-2B+A} - \sigma_I^{-B}) |f^{{}_{||_I}}{}_{{}_{\bot_I} {}_{\bot_I}}|^2(\sigma_J) \nn
\eea
where ${\cal R}_6(\sigma_J)= {\cal R}_6 (\sigma_K)|_{\sigma_I=1}$ and we refer to \cite{Andriot:2018ept} for more definitions. Finally, the source terms in the potential are here only generated from the DBI action, given as in \cite{Andriot:2017jhf} for each source. Then, one verifies that the internal metric fluctuations only arise from each parallel volume form ${\rm vol}_{||}$. One should sum all sources actions, which can be gathered in their sets as $\sum_{I} \sum_{{\rm sources}\in I}$, and each source in the set $I$ has the same ${\rm vol}_{||_I}$. The latter should be fluctuated with respect to all $\sigma_K$, and not only $\sigma_I$. This amounts to know how many parallel directions are common to two sets of sources $I$ and $J$: we denote this number by $\delta_{a_{||_I}}^{a_{||_J}} $, while the number of directions parallel to $I$ but transverse to $J$ is $p-3- \delta_{a_{||_I}}^{a_{||_J}}$. From this, we deduce the following fluctuation
\beq
{\rm vol}_{||_I} = \rho^{\frac{p-3}{2}} \sigma_I^{\frac{A(p-3)}{2}}\Pi_{J\neq I}\, \sigma_J^{\frac{(A-B)\delta_{a_{||_I}}^{a_{||_J}} + B (p-3)}{2}}  {\rm vol}_{||_I}^0 \ .
\eeq
Then, one gathers the whole sources contributions into the symbol $T_{10}^I$ as defined in \cite{Andriot:2017jhf}, and easily expresses the source terms in the potential, eventually given in \eqref{pot}.\\

From this potential, we now obtain the first and second derivatives with respect to $\sigma_I$. As we set $\sigma_{J\neq I}=1$ at the extremum, the dependence on the latter does not matter. The dependence on $\sigma_I$ is the same as in the parallel case, except for the source terms: we get for those
\beq
\sigma_I \del_{\sigma_I} \tV|_{{\rm sources} \, 0} = - \frac{g_s}{2(p+1)} \left( (A-B) ((p-3) T_{10}^I + \sum_{J\neq I} \delta_{a_{||_J}}^{a_{||_I}} T_{10}^J ) + B (p-3) T_{10} \right) \ ,
\eeq
where on top $B (p-3) T_{10} = -(A-B) (p-3) T_{10} + B (p-9) T_{10} $. We deduce, using \eqref{R6sigma3} for the curvature,
\bea
\sigma_I \del_{\sigma_I} \tV|_0 = & \ B\ {\cal R}_6 + (A - B) \left( {\cal R}_{||_I} + \delta^{ab} \del_{a_{||_I}}  f^{c_{||_I}}{}_{c_{||_I}b_{||_I}}  +  {\cal R}_{||_I}^{\bot_I} \right)  + \frac{1}{2} (A-B) |f^{{}_{||_I}}{}_{{}_{\bot_I} {}_{\bot_I}}|^2  \nn\\
& - \frac{1}{2}  \sum_n (An+B(3-n)) |H^{(n)_I}|^2 \label{delVsigma} \\
& - g_s  B\frac{p-9}{2} \frac{T_{10}}{p+1} - \frac{g_s}{2} (A-B) \left( (p-3) \frac{T_{10}^I}{p+1} + \sum_{J\neq I} \delta_{a_{||_J}}^{a_{||_I}} \frac{T_{10}^J}{p+1}   - (p-3) \frac{T_{10}}{p+1} \right) \nn\\
& -\frac{1}{2} g_s^2 \bigg(  \sum_{q=0}^{4}  \sum_n  (An+B(q-n)) |F_q^{(n)_I}|^2   \nn\\
& \phantom{+\frac{1}{2} g_s^2  \bigg(} + \frac{1}{2}  \sum_n   ( (An+B(5-n)) |F_5^{(n)_I}|^2 - (An+B(1-n)) |(*_6 F_5)^{(n)_I}|^2 ) \bigg) \ . \nn
\eea
The source terms in $(A-B)$ are new with respect to the parallel case; they vanish for $N=1$, i.e.~when there is no $J\neq I$. We rewrite the above as follows
\bea
\sigma_I \del_{\sigma_I} \tV|_0 = & \ B \bigg( {\cal R}_6 - \frac{3}{2} |H|^2 - g_s \frac{p-9}{2} \frac{T_{10}}{p+1} -\frac{1}{2} g_s^2 \big(  \sum_{q=0}^{4} q |F_q|^2 + 2 |F_5|^2  \big) \bigg) \\
+ & (A - B) \bigg( {\cal R}_{||_I} + \delta^{ab} \del_{a_{||_I}}  f^{c_{||_I}}{}_{c_{||_I}b_{||_I}}  +  {\cal R}_{||_I}^{\bot_I} + \frac{1}{2} |f^{{}_{||_I}}{}_{{}_{\bot_I} {}_{\bot_I}}|^2 \nn\\
& \qquad \quad - \frac{g_s}{2(p+1)}  \big( (p-3) (T_{10}^I - T_{10}) + \sum_{J\neq I} \delta_{a_{||_J}}^{a_{||_I}} T_{10}^J \big) \nn\\
& \qquad \quad - \frac{1}{2}  \sum_n n |H^{(n)_I}|^2 -\frac{1}{2} g_s^2 \sum_n  n \big(  \sum_{q=0}^{4}   |F_q^{(n)_I}|^2 + \frac{1}{2} (  |F_5^{(n)_I}|^2 - |(*_6 F_5)^{(n)_I}|^2 ) \big) \bigg) \ . \nn
\eea
Setting the above to zero, and using \eqref{eqdelrhoV} and \eqref{R4T10F}, we rewrite it as follows
\bea
0 =&\ \frac{3}{2} B ({\cal R}_4 + g_s^2 |F_5|^2 + 2 g_s^2 |F_6|^2) \label{eqdelsigV}\\
+ & (A - B) \bigg( {\cal R}_{||_I} + \delta^{ab} \del_{a_{||_I}}  f^{c_{||_I}}{}_{c_{||_I}b_{||_I}}  +  {\cal R}_{||_I}^{\bot_I} + \frac{1}{2} |f^{{}_{||_I}}{}_{{}_{\bot_I} {}_{\bot_I}}|^2 \nn\\
& \qquad \quad - \frac{g_s}{2(p+1)}  \big( (p-3) (T_{10}^I - T_{10}) + \sum_{J\neq I} \delta_{a_{||_J}}^{a_{||_I}} T_{10}^J \big) \nn\\
& \qquad \quad -\frac{1}{2} \sum_n  n \bigg( |H^{(n)_I}|^2 +  g_s^2 \sum_{q=0}^{4} |F_q^{(n)_I}|^2 + \frac{g_s^2}{2}  ( |F_5^{(n)_I}|^2 - |(*_6 F_5)^{(n)_I}|^2 ) \bigg) \bigg) \ . \nn
\eea
This equation is precisely the (6d integral of) the trace of the 10d Einstein equation along internal directions parallel to the sources of the set $I$, namely equations (3.18) and (3.19) of \cite{Andriot:2017jhf}; this is an important check.\footnote{This holds, as for the parallel case, up to the term in $\delta^{ab} \del_{a_{||_I}}  f^{c_{||_I}}{}_{c_{||_I}b_{||_I}}$. This term vanishes for group manifold, so we do not discuss it further here, and refer to \cite{Andriot:2018ept} about it.} As for the parallel case, we conclude on the equivalence between the 10d and 4d approaches.

We finally turn to the second derivative: the derivatives with respect to $\sigma_I$ are given by
\bea
\sigma_I^2 \del^2_{\sigma_I} \tV|_0 = & \ -\sigma_I \del_{\sigma_I} \tV|_0 -B^2\ {\cal R}_6 + (A - B)^2 |f^{{}_{\bot_I}}{}_{{}_{||_I} {}_{||_I}}|^2 \label{deldelVsigma}\\
& - (A - B) \bigg( (A+B) ( {\cal R}_{||_I} + \delta^{ab} \del_{a_{||_I}}  f^{c_{||_I}}{}_{c_{||_I}b_{||_I}} +  {\cal R}_{||_I}^{\bot_I} )  + \frac{1}{2} (3B-A)  |f^{{}_{||_I}}{}_{{}_{\bot_I} {}_{\bot_I}}|^2  \bigg) \nn\\
& + \frac{1}{2}  \sum_n ((A-B)n + 3B)^2 |H^{(n)_I}|^2 \nn\\
& - \frac{g_s}{4(p+1)} \bigg( A^2(p-3)^2 \, T_{10}^I + \sum_{J\neq I} ((A-B) \delta_{a_{||_J}}^{a_{||_I}} + B (p-3) )^2 \,  T_{10}^J  \bigg) \nn\\
& +\frac{1}{2} g_s^2 \bigg(  \sum_{q=0}^{4}  \sum_n  ((A-B)n + qB)^2 |F_q^{(n)_I}|^2   \nn\\
& \phantom{+\frac{1}{2} g_s^2  \bigg(} + \frac{1}{2}  \sum_n   ( ((A-B)n + 5B)^2 |F_5^{(n)_I}|^2 - ((A-B)n + B)^2 |(*_6 F_5)^{(n)_I}|^2 ) \bigg) \ , \nn
\eea
where the source terms could again be rewritten.

\end{appendix}

\end{document}